\documentclass[aps,pra,showpacs,twoside,10pt,floatfix,nofootinbib,longbibliography,twocolumn]{revtex4-1}
\usepackage[colorlinks=true, citecolor=red, urlcolor=blue ]{hyperref}
\usepackage{epsfig,newlfont,amssymb,amsfonts,amsmath,bm,subfigure,palatino,mathtools,amsthm,braket,soul,enumitem,color,times,comment,geometry,xcolor}
\usepackage[normalem]{ulem}
\definecolor{iitcolor}{HTML}{F7A600}
\definecolor{checkcolor}{HTML}{7541C0}

\begin{document}

\title{Estimating entanglement in 2D Heisenberg model in the strong rung-coupling limit}
\author{Chandrima B. Pushpan, Harikrishnan K. J., Prithvi Narayan, Amit Kumar Pal}
\affiliation{Department of Physics, Indian Institute of Technology Palakkad, Palakkad 678 623, India}
\date{\today}

\begin{abstract}
In this paper, we calculate entanglement in the isotropic Heisenberg model in a magnetic field on a two-dimensional rectangular zig-zag lattice in the strong rung-coupling limit, using the one-dimensional $XXZ$ model as a proxy. Focusing on the leading order in perturbation, for arbitrary size of the lattice, we show how the one-dimensional effective description emerges. We point out specific states in the low-energy sector of the two-dimensional model that are well-approximated by the one-dimensional spin-$1/2$ $XXZ$ model. We propose a systematic approach for mapping matrix-elements of operators defined on the two-dimensional model to their low-energy counterparts on the one-dimensional $XXZ$ model. We also show that partial trace-based description of entanglement in the two-dimensional model can be satisfactorily approximated using the one-dimensional $XXZ$ model as a substitute. We further show numerically that the one-dimensional $XXZ$ model performs well in estimating entanglement quantified using a measurement-based approach in the two-dimensional model  for specific choices of measured Hermitian operators.                 
\end{abstract}

\maketitle

\section{Introduction}
\label{sec:intro}

The interface of quantum information theory~\cite{nielsen2010,wilde_book} and \emph{low-dimensional} interacting quantum spin systems~\cite{Bose2001,Vasiliev2018}, with small lattice dimension with each lattice site hosting a Hilbert space of a few levels, has grown into a rich area of interdisciplinary research~\cite{Amico2008,Latorre_2009,Modi2012,laflorencie2016,DeChiara_2018,Bera_2018} in the past two decades. On one hand, these interacting quantum spin models have been identified as the natural candidates for testing and implementing quantum information and computation protocols, such as quantum state transfer~\cite{Bose2003,Burgarth2005,Burgarth2005a,Burgarth2005b,Vaucher_2005,Bose2013_chapter}, measurement-based quantum computation~\cite{raussendorf2001,raussendorf2003,briegel2009,Wei2018}, and topological quantum error corrections~\cite{kitaev2001,kitaev2006,bombin2006,bombin2007}. The motivation behind these studies has its origin in the natural occurrence of quantum states with rich quantum correlations belonging to both entanglement-separability~\cite{horodecki2009,guhne2009} and quantum information theoretic paradigms~\cite{Modi2012}, which, alongside being fundamentally important can be used as resource in several quantum tasks~\cite{horodecki2009,Modi2012}. On the other hand, these quantum correlations have provided a refreshing perspective of characterizing quantum many-body systems~\cite{Amico2008}, along with the development of tools and techniques like projected entangled pair states~\cite{Schollwock2005,Verstraete2008,Schollowck2011,Orus2014,Bridgeman_2017}, and multiscale entanglement renormalization ansatz~\cite{Verstraete2004b,Vidal2007,Vidal2008,Rizzi2008,Aguado2008,Cincio2008,Evenbly2009}. Current experimental advances allowing implementation and manipulation of these quantum spin models as well as quantum protocols designed on these models using trapped ions \cite{Porras2004,Leibfried2005,monz2011,Korenblit_2012,Bohnet2016}, superconducting qubits\cite{barends2014,Yariv2020}, nuclear magnetic resonance \cite{Vandersypen2005,Negrevergne2006}, solid-state systems\cite{Schechter2008,Bradley2019}, and ultra-cold atoms \cite{Greiner2002,Duan2003,Bloch_2005,Bloch2008,Struck2013} have also provided a major boost to these studies.  

Among a plethora of low-dimensional quantum spin models, two-dimensional (2D) lattice models have always been specially challenging due to the faster growth of  Hilbert space dimension with increasing number of spins in the system, compared to their one-dimensional (1D) counterparts. One such model is the Heisenberg model~\cite{Heisenberg1928,Okwamoto1984,Aplesnin1999,Zheng1999,Costa2003,Cuccoli2006,Ju2012,Verresen2018,Sariyer2019} in a magnetic field on a rectangular lattice of $NL$ sites, having respectively $N$ and $L$ lattice sites in the horizontal and vertical directions, where each lattice site hosts a spin-$1/2$ particle. A number of recent studies~\cite{Weihong1999,Kallin2011,Song2011,Lima2020} have been carried out to understand the entanglement properties of the model. Particular attention has been drawn towards quasi-1D models~\cite{Ercolessi2003,Ivanov2009} like quantum spin ladders~\cite{Dagotto1996,Batchelor2003,Batchelor2003a,Batchelor2007} with number of lattice sites in the horizontal direction being far greater than the same in the vertical direction ($N\gg L$). As natural extensions of the 1D models while going towards 2D, quantum spin ladders with $L=2$ has been investigated from the perspective of quantum state transfer~\cite{Li2005,Almeida2019}. Moreover, entanglement~\cite{Song2006,Chen2006,Dhar_2011,Ren2011,Lauchli2012,Dhar2013,Santos2016,Singharoy2017,Almeida2019} and fidelity~\cite{Ren2011,Li_2017} have been investigated in these ladder models from the perspective of  phases  characterizations. While most of these studies have concentrated on models with spin-$1/2$ particles, entanglement properties of quantum spin ladders with higher spin quantum numbers~\cite{Schliemann2012} have also been explored.    

In the limit where the coupling strength $J_\perp$ along the rungs of the ladder are much larger compared to other spin-exchange interactions present in the system, the rung behaves like a single degree of freedom, and the model becomes effectively 1D. By further tuning magnetic field strength, for low values of $L$ ($L=2,3$), \cite{kawano1997,totsuka1998,Tonegawa1998,Mila1998,Chaboussant1998,Tandon1999} show that the isotropic Heisenberg quantum spin-$1/2$ ladders map to an 1D $XXZ$ model~\cite{fisher1964,giamarchi2004,Mila_2000,Franchini2017}. This mapping has been used to study quantum phase transitions in the case of  antiferromagentic spin-$1/2$ ladders with $L=2$~\cite{totsuka1998,Tonegawa1998,Mila1998,Chaboussant1998,Tandon1999,Tribedi2009} and $3$~\cite{kawano1997,Tandon1999}, using magnetization properties~\cite{totsuka1998,Tonegawa1998,Chaboussant1998,Tandon1999}, and entanglement~\cite{Tribedi2009}  (see also~\cite{Chen2007} for the mapping in the case of a mixed-spin variant of quantum spin ladders with $L=2$). While most of these studies have focused on local observables,  non-local correlations such as entanglement in the low-energy sector of the spin-$1/2$ Heisenberg model on a quasi-1D, or 2D lattice, using the effective 1D model as a proxy, is yet to be explored, to the best of our knowledge. 

\begin{figure}
    \centering
    \includegraphics[width=0.8\linewidth]{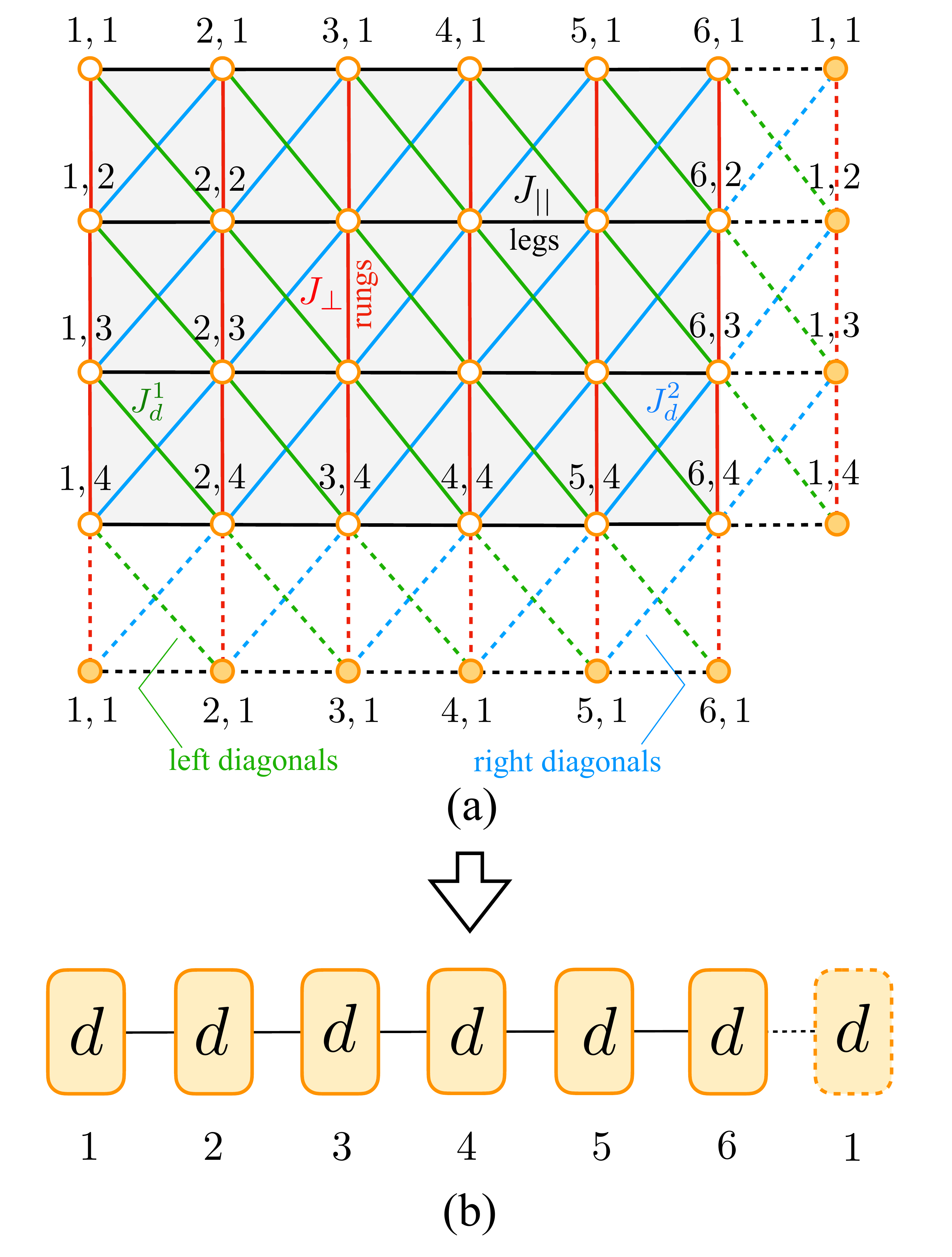}
    \caption{(a) A two-dimensional zig-zag lattice of $6$ rungs and $4$ legs holding $6\times 4$ spin-$\frac{1}{2}$ particles. The spin-exchange interaction strength along the rungs and the legs are $J_{\perp}$ and $J_{||}$, respectively, while the spins interact diagonally with the interaction strengths $J_d^1$ and $J_d^2$ along the left and the right diagonals respectively. The periodic boundary condition along the rungs, legs, and diagonals are represented by the dashed links to the edge spins of the lattice. Each spin in the lattice is subjected to a magnetic field of strength $h$ in the $z$-direction. (b) Using degenerate perturbation theory in the low-energy subspace in the strong rung-coupling limit ($J_\perp\gg J_{||},J_d^1,J_d^2$), for specific points in the parameter space of $J_{\perp}$ and $h$, the model can be mapped to a effective 1D Hamiltonian (see Eq.~(\ref{eq:first_order_effective_Hamiltonian_general})), where each lattice site hosts a $d$-dimensional Hilbert space corresponding to each rung of the system. }
    \label{fig:fig1_schematics}
\end{figure}

In this paper, we specifically ask \emph{whether entanglement properties in the 1D effective model can approximate the corresponding entanglement in all states of the low-energy manifold of the  isotropic Heisenberg model in a magnetic field, defined on a quasi-1D, or 2D rectangular zig-zag lattice (see Fig.~\ref{fig:fig1_schematics}(a))}.
Towards approaching this question, we lay the ground by determining the 1D effective model in a system size-independent fashion, thereby bringing the specific models, studied so far in a case-by-case basis~\cite{kawano1997,totsuka1998,Tonegawa1998,Mila1998,Chaboussant1998,Tandon1999,Tribedi2009}, under the same umbrella. As mentioned above, we particularly focus on tuning the magnetic  field such that the ground-state degeneracy of the model in the strong rung-coupling limit is two-fold. We prove that the 1D effective Hamiltonian  is rotationally invariant about the $z$-axis to any order in perturbation theory. We focus on the leading order in perturbation, which turns out to be an 1D spin-$1/2$ $XXZ$ model,  which can be dealt with relatively easily due to the computational advantage obtained from the reduction in the Hilbert space, and  considerable information  is available about the model in literature~\cite{fisher1964,giamarchi2004,Mila_2000,Franchini2017}. We analytically derive the effective coupling constants of the 1D $XXZ$ model as functions of the coupling constants in the original model for arbitrary values of $N$ and $L$ and for different boundary conditions on the 2D lattice in the horizontal and vertical directions. We also point out a set of typical states including the thermal state and the time-evolved state constituted from the low-energy sector of the model for which the effective 1D theory works, up to $\mathcal{O}(J_\perp^{-1})$, in estimating the states.  We give a systematic prescription for mapping matrix-elements of operators among these states to their 1d effective counterparts. We also show that the entanglement  in low-energy states of the quasi-1D or 2D model is satisfactorily proxied by the entanglement in the corresponding mapped state in the 1D model. We illustrate these results using typical observables and  entanglement measures computed using partial trace-based methods. We also investigate measurement-based entanglement in the quasi-1D and 2D model, for which analytical treatment is difficult.  However, for certain choices of measured Hermitian operators, our numerical analysis shows that the 1D effective $XXZ$ model can mimic the results in the quasi-1D or 2D model.

The rest of the paper is organized as follows. In Sec.~\ref{sec:methodology}, we discuss the 2D Heisenberg model, the low-energy 1D effective Hamiltonian, and its symmetry, and point out typical states where the 1D effective theory performs satisfactorily. The technical details of working out the 1D effective $XXZ$ model are included in Sec.~\ref{sec:details}. Sec.~\ref{sec:mapping_observables} deals with the performance of the 1D effective theory in computing matrix elements of Hermitian operators, while the application of the 1D effective theory in quantifying entanglement using partial trace-based and measurement-based approaches are included in Sec.~\ref{sec:entanglement}. Sec.~\ref{sec:outlook} contains the concluding remarks and outlook.

\section{Low-energy effective Hamiltonian}
\label{sec:methodology}

In this section, we introduce the Isotropic Heisenberg model on the rectangular zig-zag lattice, and discuss mapping it to an effective one-dimensional (1D) lattice model in the low-energy subspace. 

\subsection{Isotropic Heisenberg model on a 2D zig-zag lattice}
\label{subsec:system_main}

Let us consider an  $N\times L$ lattice with $L$ $(N)$ lattice sites along the vertical (horizontal) direction, each lattice site hosting a spin-$\frac{1}{2}$ particle (see Fig.~\ref{fig:fig1_schematics}(a)), such that a $2^{NL}$-dimensional Hilbert space $\mathcal{H}$ is associated to the system. We call the vertical (horizontal) lines in the lattice as the \emph{rungs} (\emph{legs}), where $N$ ($L$) represents the number of rungs (legs). While $N=L$ represents a zig-zag square lattice, $N\gg L$  represents a zig-zag \emph{ladder} that is intermediate between the one- and two-dimensional lattices, and is therefore called a \emph{quasi}-1D lattice model~\cite{Batchelor2007}. Isotropic Heisenberg interactions~\cite{fisher1964,giamarchi2004,Mila_2000,Franchini2017} are present between the spins along the legs, rungs, as well as the diagonals, and an external magnetic field applies to all spins along the $z$ direction. The spin system is represented by the Hamiltonian~\cite{Batchelor2007}
\begin{eqnarray}
H/J_\perp=H_R+H_I,
\label{eq:system_Hamiltonian}
\end{eqnarray}
with 
\begin{eqnarray}
\label{eq:system_Hamiltonian_rung}
H_R&=&\frac{1}{4}\sum_{i=1}^N\sum_{j=1}^{L}\vec{\sigma}_{i,j}.\vec{\sigma}_{i,j+1}-\frac{h}{2J_\perp}\sum_{i=1}^N\sum_{j=1}^L\sigma^z_{i,j}-NE_g,\nonumber\\
\end{eqnarray}
and
\begin{eqnarray}
H_I&=&\frac{J_{||}}{4J_\perp}\sum_{i=1}^N\sum_{j=1}^L\vec{\sigma}_{i,j}.\vec{\sigma}_{i+1,j}+\frac{J_{d}^1}{4J_\perp}\sum_{i=1}^N\sum_{j=1}^{L}\vec{\sigma}_{i,j+1}.\vec{\sigma}_{i+1,j}\nonumber\\
&&+\frac{J_{d}^2}{4J_\perp}\sum_{i=1}^N\sum_{j=1}^{L}\vec{\sigma}_{i,j}.\vec{\sigma}_{i+1,j+1}.
\label{eq:system_Hamiltonian_leg}
\end{eqnarray}
Here, $J_\perp$ ($J_{||}$) is the strength of the spin-exchange interactions along the rungs (along the legs), $J_d^1$ and $J_d^2$ are the diagonal interaction strengths, $h$ is the strength of the magnetic field, $\vec{\sigma}_{i,j}\equiv\{\sigma^x_{i,j},\sigma^y_{i,j},\sigma^z_{i,j}\}$ are the Pauli operators on the lattice-site denoted by the subscripts $i,j$, with $i$ ($j$) being the rung (leg) index running from 1 to $N$ ($L$), and $E_g$ is a constant we will choose in section \ref{subsec:leh}.  We use the computational basis $\{\ket{0},\ket{1}\}$ to write the Pauli matrices, where $\sigma^z\ket{0}=\ket{0}$, $\sigma^z\ket{1}=-\ket{1}$, such that $\ket{0}\equiv\ket{\uparrow}$, and $\ket{1}\equiv\ket{\downarrow}$.

\subsection{Effective Hamiltonian in low-energy subspace}
\label{subsec:leh}
 
We now discuss the construction of a low-energy effective 1D Hamiltonian (LEH)~\cite{Mila2011} corresponding to the spin models described in Sec.~\ref{subsec:system_main}.  We start with the limit $J_{||}=J_d^1=J_d^2=0$, where the model consists of $N$ non-interacting rungs, each of which corresponds to a $2^L$-dimensional Hilbert space. The antiferromagnetic isotropic Heisenberg Hamiltonian of each rung is given by 
\begin{eqnarray}
\label{eq:system_Hamiltonian_one_rung}
H_{R_i}=\frac{1}{4}\sum_{j=1}^{L}\vec{\sigma}_{i,j}.\vec{\sigma}_{i,j+1}-\frac{h}{2J_\perp}\sum_{j=1}^L\sigma_{i,j}^z-E_g,
\end{eqnarray}
with $H_R=\sum_{i=1}^NH_{R_i}$. Let us assume that for a given value of $h$, the spectrum of $H_{R_i}$ is given by $\{E_{k_i}^{(i)},\ket{\psi_{k_i}^{(i)}}\}$, where  $k_i\in\{0,1,\cdots,2^L-1\}$ $\forall i\in\{1,2,\cdots,N\}$, such that 
\begin{eqnarray}
H_{R_i}\ket{\psi_{k_i}^{(i)}}=E_{k_i}^{(i)}\ket{\psi_{k_i}^{(i)}},
\end{eqnarray}
with $k_i=0$ representing the ground state.  A $d$-fold degeneracy in the ground states can be imposed via tuning $h$ to specific values, denoted by $h^\prime$ 
where typically $h^\prime\sim J_\perp$. At this point, we choose $E_g$ such that ground state energy of $H_R$ with $h=h^\prime$ vanishes. For a fixed $L$, there may be multiple values of $h^\prime$ resulting in a $d$-fold degenerate ground state. For each $h^\prime$, we relabel the states $\{\ket{\psi_{k_i}^{(i)}}\}$ such that
\begin{eqnarray}
E_{k_i}^{(i)}&=&0\;\forall\; k_i=0,1,2,\dots,d-1,\nonumber\\
E_{k_i}^{(i)}&>&0\;\forall\; k_i=d,d+2,\dots,2^L-1,
\end{eqnarray}
and denote the $d$-fold degenerate ground states by $\{\ket{\psi_0^{(i)}},\cdots,\ket{\psi_{d-1}^{(i)}}\}$. For each $h^\prime$, the ground-state manifold $\{\ket{\Psi_l}\}$ of $H_R$ is $d^N$-fold degenerate, constituting the low-energy subspace $\mathcal{S}\subset\mathcal{H}$, with $\ket{\Psi_l}$ having the form
\begin{eqnarray}
\ket{\Psi_l}=\bigotimes_{i=1}^N\ket{\psi_{k_i}^{(i)}}.
\label{eq:ground_state_manifold}
\end{eqnarray}
Here, $k_i=0,1,\cdots,d-1$, and $l=0,1,2,\cdots,d^N-1$ labels the ground state manifold\footnote{Note that $l$ can be identified as the decimal equivalent of the string $k_1k_2k_3\cdots k_N$ in base $d$.}. We will find it useful to split the Hilbert space into two subspaces, one is spanned by $\ket{ \Psi_l  }$ (which we will term as {\it low energy sector}) and the other (which we will term as {\it high energy sector}) orthogonal to this space. We can also define a projector onto the low energy sector as
\begin{equation}
    P_g = \sum_l \ket{\Psi_l} \bra{\Psi_l}
\end{equation}
and the projector on the high energy sector is $P_e \equiv I - P_g$.

Let us now rewrite the system Hamiltonian $H$ (Eq.~(\ref{eq:system_Hamiltonian})) as 
\begin{eqnarray}
\frac{1}{J_\perp}H=H_0+\frac{1}{J_\perp}H^\prime,
\label{eq:system_Hamiltonian_rewrite}
\end{eqnarray}
where 
\begin{eqnarray}
H_0&=&\frac{1}{4}\sum_{i=1}^N\sum_{j=1}^{L}\vec{\sigma}_{i,j}.\vec{\sigma}_{i,j+1}-\frac{h^\prime}{2J_\perp}\sum_{i=1}^N\sum_{j=1}^L\sigma^z_{i,j}-NE_g,\nonumber\\
\label{eq:system_Hamiltonian_unperturbed}
\end{eqnarray}
which is $H_R$ at $h=h^\prime$, and 
\begin{eqnarray}
H^\prime &=& \frac{J_{||}}{4}\sum_{i=1}^N\sum_{j=1}^L\vec{\sigma}_{i,j}.\vec{\sigma}_{i+1,j}+\frac{J_{d}^1}{4}\sum_{i=1}^N\sum_{j=1}^{L}\vec{\sigma}_{i,j+1}.\vec{\sigma}_{i+1,j}\nonumber\\
&+&\frac{J_{d}^2}{4}\sum_{i=1}^N\sum_{j=1}^{L}\vec{\sigma}_{i,j}.\vec{\sigma}_{i+1,j+1}-\frac{\Delta h}{2}\sum_{i=1}^N\sum_{j=1}^L\sigma^z_{i,j},
\label{eq:system_Hamiltonian_perturbation}
\end{eqnarray}  
with $\Delta h=h-h^\prime$. Clearly, ground states of $H_0$ have a $d^N$-fold degeneracy (see Eq.~(\ref{eq:ground_state_manifold})), which is lifted by the perturbation $H^\prime$.
For $\Delta h, J_{||},  J_d^{1,2} \ll  J_\perp$, this leads to an effective Hamiltonian operating in the low-energy subspace $\mathcal{S}$, where energy-eigenvalues of the $n$-th order $(n = 1, 2, ...)$ effective Hamiltonian in $\mathcal{S}$ provides the $n$-th order energy corrections to the unperturbed states of low energy. In this paper, we focus only on the first-order ($n=1$) effective Hamiltonians, which can be obtained as 
\begin{eqnarray}
\label{eq:first_order_effective_Hamiltonian_general}
\tilde{H} &=& \sum_{l,l^\prime=0}^{d^N-1}\bra{\Psi_{l}}H^\prime\ket{\Psi_{l^\prime}}\ket{\Psi_{l}} \bra{\Psi_{l^\prime}}.
\end{eqnarray}
Note here that the first-order effective Hamiltonian (\ref{eq:first_order_effective_Hamiltonian_general}) represents a system of one lattice dimension with $N$-sites, where each lattice site has $d$ degrees of freedom.

In this paper, we focus on a subset of cases with $d=2$ by choosing $h'$ appropriately. The effective degrees of freedom in this case is like a spin half system. The effective Hamiltonian will be built out of operators acting on this low energy spin-$1/2$ degree of freedom. While we can use the definition (\ref{eq:first_order_effective_Hamiltonian_general}) to find the effective Hamiltonian, we find it more instructive to work out the symmetries of the system which will then constrain the structure of the effective Hamiltonian. The detailed structure of the effective Hamiltonian can be subsequently worked out. We denote the doubly-degenerate ground states of $H_{R_i}$ at $h=h^\prime$ by $\ket{\psi_0^{(i)}}$ and $\ket{\psi_1^{(i)}}$. Since $[Z_i,H_{R_i}]=0$, with $Z_i=\sum_{j=1}^L\sigma^z_{i,j}$ defined on the rung $i$,  
\begin{eqnarray}
Z_i\ket{\psi_{k}^{(i)}}=m_{k}\ket{\psi_{k}^{(i)}},
\label{eq:rung_rotation_generator_eigenvalue_equation}
\end{eqnarray}
$k=0,1$, where $m_k$ are the eigenvalues of $Z_i$ corresponding to the eigenvectors $\ket{\psi_{k}^{(i)}}$. Note here that $Z_i$ is the generator of the $z$-rotation on the Hilbert space of the $i$th rung. We now propose the following.

\noindent$\blacksquare$ \textbf{Proposition 1.} \emph{If $m_0\neq m_1$, $\tilde{H}$ is $z$-rotationally invariant.}

\begin{proof}

Consider the operator
\begin{eqnarray}
\mathcal{Z} &=& \sum_{i=1}^N J_i^z,
\end{eqnarray}
with 
\begin{eqnarray}
J_i^z=a+bZ_i, 
\end{eqnarray} 
where $\mathcal{Z}$ is a symmetry of the original system-Hamiltonian $H$, i.e., 
\begin{eqnarray}
[\mathcal{Z},H_0]=[\mathcal{Z},H^\prime]=0, 
\label{eq:commutation}
\end{eqnarray}
and $a$ and $b$ are real constants that can be chosen according to convenience. Since  $m_0\neq m_1$, we choose $a=(m_1+m_0)/(m_1-m_0)$ and $b=2/(m_0-m_1)$ such that 
\begin{eqnarray}\label{eq:Action of J_z}
J_i^z\ket{\psi_0^{(i)}}=\ket{\psi_0^{(i)}},\quad J_i^z\ket{\psi_1^{(i)}}=-\ket{\psi_1^{(i)}}.   
\end{eqnarray}
Therefore, the action of $J^z_i$ on the rung-subspace spanned by $\ket{\psi_{0(1)}^{(i)}}$ is similar to that of an \emph{effective} Pauli-$z$ operator\footnote{Note that the assumption $m_0\neq m_1$ is crucial for the $J_i^z$s to mimic the action of Pauli-$z$ operators}, which can be defined as  
\begin{eqnarray}
\label{eq:epauliz}
\tau^z_i &=& \ket{\psi_0^{(i)}}\bra{\psi_0^{(i)}}-\ket{\psi_1^{(i)}}\bra{\psi_1^{(i)}}.
\end{eqnarray}

Let us further define the operator $\eta^z=\sum_{i=1}^N\tau^z_i$, and note that 
\begin{eqnarray}\label{eq:Action of Z on ground state manifold}
\eta^z\ket{\Psi_l}=\mathcal{Z}\ket{\Psi_l}=m_l\ket{\Psi_l},
\end{eqnarray}
where $m_l=\sum_{i=1}^{N}(-1)^{k_i}$ (see Eq.~(\ref{eq:ground_state_manifold})). 
Since $[ \eta_z, H' ]= 0$, $\eta_z$ can't change the eigenvalue under $H'$ and hence $\left\langle\Psi_l\right|H^\prime\left|\Psi_{l^\prime}\right\rangle =  0$ if $m_l \ne m_{l^\prime}$. 
Since this vanishes for $m_l \ne m_l'$, we have  
\begin{eqnarray}
[\eta^z,\tilde{H}] &=& \sum_{l,l^\prime=0}^{2^N-1}\left(m_{l}-m_{l^\prime}\right) \ket{\Psi_l}\bra{\Psi_{l}}H^\prime\ket{\Psi_{l^\prime}}\bra{\Psi_{l^\prime}}\nonumber\\
&=& 0,
\end{eqnarray}
implying that $\eta^z$ is a symmetry of the effective 1D Hamiltonian $\tilde{H}$. 
\end{proof} 

Since the first order effective Hamiltonian is guaranteed to be at most nearest-neighbour in the effective spins, the following corollary can be written. 
 
\noindent$\diamond$ \textbf{Corollary 1.1} \emph{The most general form of $z$-rotationally invariant $\tilde{H}$ is a nearest-neighbor $XXZ$ model}~\cite{giamarchi2004} \emph{in a magnetic field,  given by}   
\begin{eqnarray}
\tilde{H}&=&\sum_{i=1}^N\left[ \tilde{J}_{xy}\left(\tau^x_i\tau^x_{i+1}+\tau^y_i\tau^y_{i+1}\right)+\tilde{J}_{zz}\tau^z_i\tau^{z}_{i+1}\right]\nonumber\\
\affiliation{}&&+\tilde{h}\sum_{i=1}^N \tau^z_i + C,
\label{eq:1d_general}
\end{eqnarray}
\emph{where $C$ is an irrelevant additive constant.} 

\noindent Here, we have defined 
\begin{eqnarray}
\label{eq:epaulix}
\tau^x_i &=& \ket{\psi_0^{(i)}}\bra{\psi_1^{(i)}}+\ket{\psi_1^{(i)}}\bra{\psi_0^{(i)}} ,\nonumber\\
\label{eq:epauliy}
\tau^y_i &=& -\text{i}\left[\ket{\psi_0^{(i)}}\bra{\psi_1^{(i)}}-\ket{\psi_1^{(i)}}\bra{\psi_0^{(i)}}\right]
\end{eqnarray} 
as the effective Pauli-$x$ and $y$ operators defined on the rung subspace spanned by $\ket{\psi_{0,1}^{(i)}}$ (see Eq.~(\ref{eq:epauliz})).  The coefficients $\tilde{J}_{xy}$, $\tilde{J}_{z}$, and $\tilde{h}$ can be determined as functions of the couplings in the perturbation Hamiltonian $H^\prime$, namely, $J_{||}$, $J_d^{1,2}$, and $\Delta h$.  The exact forms of the functions depend on the forms of the doubly-degenerate ground states $\ket{\psi_{0,1}^{(i)}}$, and subsequently on the specific point $h=h^\prime$ where the perturbation calculation is carried out. 

\noindent\textbf{Note 1.} In the low-energy sector, it is sufficient to consider the effective Hamiltonian (\ref{eq:1d_general}) as a model for $N$ spin-$1/2$ particles, where the Pauli matrices on the Hilbert space of the spin-$1/2$ particle are given by $\tau^{x,y,z}$. In this picture, we relabel $\ket{\psi^{(i)}_{0,1}}$ as
\begin{eqnarray}
\label{eq:XXZ_states}
\ket{\mathbf{0}_i}=\ket{\psi^{(i)}_0}, \ket{\mathbf{1}_i}=\ket{\psi^{(i)}_1},
\end{eqnarray}
where $\{\ket{\mathbf{0}_i},\ket{\mathbf{1}_i}\}$ form $\tau^z$ eigenbasis of the spin-$1/2$ particle at the $i$th site of the 1D lattice of size $N$. Eigenstates of the 1D model, $\ket{\tilde{\Phi}_l}$, can be written in terms of the $\{\ket{\mathbf{0}_i},\ket{\mathbf{1}_i}\}$ basis of the spins in the system. 

\noindent\textbf{Note 2.} A couple of comments about the notations used in the rest of the paper is in order here. 
\begin{enumerate}
\item[(a)] Any state $\ket{ \tilde \Phi } $ satisfying $P_g \ket{ \tilde \Phi } = \ket{ \tilde \Phi }$ is essentially a state in the 1D $XXZ$ model, which can be written in terms of $\ket{\mathbf{0}}$ and $\ket{\mathbf{1}}$. We will typically put $\tilde \ $ on such states to remind us that it is  an $XXZ$ state. Note that the projection of a state $\ket{\Phi}$ onto the low energy sector is $P_g \ket{\Phi}$.

\item[(b)] Similarly, any operator $\tilde A$ satisfying $P_g \tilde A P_g = \tilde A$ is essentially an operator defined on the Hilbert space of the 1D $XXZ$ model, which can be written in terms of $\tau^{x,y,z}$.  We will typically put $\tilde \ $ on such operator to remind us that it is an $XXZ$ operator. This also applies to the state $\rho$, which is a Hermitian operator. Note that the projection of an operator  $A$ onto the low energy sector is $P_gA P_g$.
\end{enumerate}

\subsection{Merit of the mapping}
\label{subsec:merit_of_mapping}

We now ask how well the low-energy spectrum of the 2D model is captured by the 1D effective Hamiltonian $\tilde H$. In Sec.~\ref{subsec:leh}, we see that for $h = h^\prime$ and $\Delta h = J_{||}=  J_d^{1,2} =  0$, the Hilbert space of the system naturally splits into two subspaces - (a) the low-energy subspace $\mathcal{S}$, spanned by the orthonormal degenerate states $\{\ket{\Psi_l}\}$, where each of the state $\ket{\Psi_l}$ has zero energy w.r.t. $H_0$, and (b) the excited subspace $\mathcal{S}^\prime$, spanned by the excited eigenstates $\{\ket{\Psi^\prime_\alpha}\}$ of $H_0$ with energy $E_\alpha^\prime\geq \mathcal{O}(J_\perp)$, where  
$\langle\Psi^\prime_\alpha|\Psi_l\rangle=0\;\forall l,\alpha$. The total state-space of the system is given by $\mathcal{S}\oplus\mathcal{S}^\prime$. We are now interested in the spectrum of the system slightly away from the point $h=h^\prime$ in the space of coupling constants i.e $\Delta h, J_{||},  J_d^{1,2} \ll  J_\perp$. Perturbation theory in the parameter ${1/J_\perp}$ gives the following structure for the energy eigenvalues and eigenvectors of the 2D model: 
\begin{eqnarray}
\label{eq:State_perturbative_corrections}
\ket{\Phi_l} &=& \ket{\Phi_l^0}+J_\perp^{-1} \ket{\Phi_l^{1}}+J_\perp^{-2} \ket{\Phi_l^2}+\dots \\
E_l &=&  E_l^{(1)} +  J_\perp^{-1} E_l^{(2)}+ \dots 
\end{eqnarray}
where $\ket{\Phi_l^0}$ and $E_l^{(1)}$ are respectively the \emph{non-degenerate}\footnote{Degenerate eigenstates of $\tilde{H}$ would result in degeneracy in the energy levels of the 2D model, which can be lifted by second, or higher order perturbation theory.} eigenvectors and eigenvalues of $\tilde{H}$. This makes it explicit that the low lying spectrum of the 2D Hamiltonian are very close (upto ${1/J_\perp}$ corrections) to the $XXZ$ spectrum. To see that the eigenstates also match, note that the first order perturbation theory gives the first-order state corrections $\ket{\Phi^1_l}$ to have overlaps with the excited state manifold $\{\ket{\Psi^\prime_\alpha}\}$ having energy of $\mathcal{O}(J_\perp)$, implying that $\ket{\Phi^1_l}$ is a small correction to $\ket{\Phi_l}$ in  Eq.~(\ref{eq:State_perturbative_corrections}) due to the suppression factor $J_\perp^{-1}$.  
One may quantify the small corrections via the \emph{distance} between  $\ket{\Phi_l}$ and $\ket{\Phi_l^0}$, which can be quantified by a distance metric, eg. the \emph{trace-distance}~\cite{wilde_book} (see Appendix~\ref{app:distance} for definition).

In the following, we describe four instances where the mapping described in Sec.~\ref{subsec:leh} is particularly useful. 

    \subsubsection{States made of non-degenerate eigenstates}
    \label{subsubsec:non_degenerate_pure_states}
    
    A state in  the Hilbert space of the 2D model, which can be written in terms of the non-degenerate eigenstates of $H$ as
    \begin{eqnarray}    \rho=\sum_{l,l^\prime}c_{l,l^\prime}\ket{\Phi_l}\bra{\Phi_{l^\prime}},
    \label{eq:2d_density_matrix}
    \end{eqnarray}
    with $\sum_l|c_{l,l}|^2=1$, can be approximated as
    \begin{eqnarray}
    \rho\approx \sum_{l,l^\prime}c_{l,l^\prime}\ket{\Phi_l^0}\bra{\Phi_{l^\prime}^0}=\tilde{\rho}, 
    \end{eqnarray}
    as long as the overall accumulated error due to the first order corrections in $\ket{\Phi_l}\bra{\Phi_{l^\prime}}$ is $\leq\mathcal{O}(J_\perp^{-1})$ \footnote{This can be ensured by keeping the number of terms in $\rho$ small.}. We remind ourselves here that $\tilde{\rho}$ is a density matrix entirely in the low energy sector (see \textbf{Note 2}, Sec.~\ref{subsec:leh}).   As an example, consider the state 
    \begin{eqnarray}
    \rho &=& \frac{1}{2}\Big[\ket{\Phi_0}\bra{\Phi_0}+\ket{\Phi_{2^N-1}}\bra{\Phi_{2^N-1}}\nonumber\\
    &+&\ket{\Phi_0}\bra{\Phi_{2^N-1}}+\ket{\Phi_{2^N-1}}\bra{\Phi_{0}}\Big]
    \end{eqnarray}    
    of the 2D model, where $\ket{\Phi_0}$ and $\ket{\Phi_{2^N-1}}$ are non-degenerate eigenstates of $H$\footnote{Here we work in a subspace of the space of coupling constants in $H$ where $\ket{\Phi_0}$ and $\ket{\Phi_{2^N-1}}$ are non-degenerate.}. This state can be approximated as 
    \begin{eqnarray}
    \rho\approx\tilde{\rho} &=& \frac{1}{2}\Big[\ket{\Phi_0^0}\bra{\Phi_0^0}+\ket{\Phi_{2^N-1}^0}\bra{\Phi_{2^N-1}^0}\nonumber\\ &+&\ket{\Phi_0^0}\bra{\Phi_{2^N-1}^0}+\ket{\Phi_{2^N-1}^0}\bra{\Phi_{0}^0}\Big]
    \end{eqnarray} 
    up to the correction $\mathcal{O}(J_\perp^{-1})$. Since $\ket{\Phi_0^0}=\otimes_i\ket{\psi_0^{(i)}}$, and $\ket{\Phi_{2^N-1}^0}=\otimes_{i}\ket{\psi_1^{(i)}}$, $\tilde{\rho}$ can be written as $\tilde{\rho}=\ket{\tilde{\Phi}_{\text{GHZ}}}\bra{\tilde{\Phi}_{\text{GHZ}}}$ in the space of the 1D effective model, where 
    \begin{eqnarray}
    \ket{\tilde{\Phi}_{\text{GHZ}}} =\frac{1}{\sqrt{2}}\left[\ket{\mathbf{0}^{\otimes N}}+\ket{\mathbf{1}^{\otimes N}}\right]
    \end{eqnarray} 
    is the $N$-qubit GHZ state~\cite{greenberger1989,zeilinger1992}.  
    Note that for ferromagnetic couplings $\tilde{J}_{xy}, \tilde{J}_{zz}<0$ and for $\tilde{h}>0$, $\ket{\textbf{0}}^{\otimes N}$ and $\ket{\mathbf{1}}^{\otimes N}$ represent the ground state and the highest energy eigenstate of $\tilde{H}$ respectively. 
    
    \subsubsection{Equal mixture of degenerate states} 
    \label{subsubsec:equal_mixture}
    
    Let us now assume that the energy level $E_{l=l^\prime}^{(1)}$ of $\tilde{H}$ is $n$-fold degenerate, where the eigenstates of $\tilde{H}$ corresponding to $E_{l^\prime}^{(1)}$ are $\ket{\Phi_{l^\prime,m}^0}$, with the index $m$ running from $1$ to $n$.  This leads to an $n$-fold degeneracy of the energy level $E_{l^\prime}$ of $H$ with degenerate eigenstates $\ket{\Phi_{l^\prime,m}}$ in first order perturbation theory. Although  $\ket{\Phi_{l^\prime,m}^0}$  does not approximate the states $\ket{\Phi_{l^\prime,m}}$ up to small corrections, the total contribution of the degenerate manifold $\{\ket{\Phi_{l^\prime,m}}\}$ in the density matrix at the low-energy sector is approximated by the contribution of the $\{\ket{\Phi_{l^\prime,m}^0}\}$. More precisely, the contribution is represented by an equal mixture of the degenerate states $\{\ket{\Phi_{l^\prime,m}}\}$ in the Hilbert space of the 2D model, given by 
    \begin{eqnarray}
    \rho=\frac{1}{n}\sum_{m=1}^n\ket{\Phi_{l^\prime,m}}\bra{\ket{\Phi_{l^\prime,m}}},
    \end{eqnarray} 
    which is approximated, up to corrections of $\mathcal{O}(J_\perp^{-1})$, by an equal mixture of the corresponding degenerate states $\{\ket{\Phi_{l^\prime,m}^0}\}$, i.e., 
    \begin{eqnarray}
    \rho\approx\tilde{\rho}=\frac{1}{n}\sum_{m=1}^n\ket{\Phi_{l^\prime,m}^0}\bra{\ket{\Phi_{l^\prime,m}^0}}. 
    \end{eqnarray} 
    
    \subsubsection{Thermal states} 
    \label{subsubsec:thermal_states}
    
    Consider the thermal state,
    \begin{eqnarray}
    \rho_T=\frac{\sum_\alpha\text{e}^{-\beta E_\alpha}\ket{\Phi_\alpha}\bra{\Phi_\alpha}}{\sum_\alpha\text{e}^{-\beta E_\alpha}},
    \end{eqnarray}
    of the 2D system at an absolute temperature $k_BT$, with $k_B$ being the Boltzmann constant, and $\beta=1/k_BT$, where $\{\ket{\Phi_\alpha}\}$ is the full spectrum of $H$. For low temperatures such that $k_BT\ll \mathcal{O}(J_\perp)$ (or, $1\ll \beta J_\perp$), the state is well-approximated by the low-energy spectrum as
    \begin{eqnarray}
    \rho_T&\approx&\frac{\sum_{l =0}^{2^N-1}\text{e}^{-\beta E_l}\ket{\Phi_l}\bra{\Phi_l}}{\sum_{l=0}^{2^N-1}\text{e}^{-\beta E_l}}\nonumber\\ &\approx& \frac{\sum_{l=0}^{2^N-1}\text{e}^{-\beta E_l^{(1)}}\ket{\Phi_l^0}\bra{\Phi_l^0}}{\sum_{l=0}^{2^N-1}\text{e}^{-\beta E_l^{(1)}}}\nonumber\\ &=&\tilde{\rho},
    \label{eq:thermal_state_H}
    \end{eqnarray} 
    up to correction of $\mathcal{O}(1/J_\perp)$. 
    In the first approximation above, we have truncated the sum to states with $E \ll {\cal O}(J_\perp)$ and in the second, we have approximated both the states and the energies by their $XXZ$ counterparts.

    \subsubsection{Closed system evolution} \label{subsubsec:time_evolution}
    
    The mapping of the 2D model to the effective 1D $XXZ$ model also allows one to investigate the time-evolution $\rho(t)$ of the closed system as long as the initial state $\rho(0)$ is of the type described in Eq.~(\ref{eq:2d_density_matrix}). Such a state in the Hilbert space of the 2D model evolves as
    \begin{eqnarray}
    \rho(t)&=&\sum_{l,l^\prime}c_lc_{l^\prime}^\star\text{e}^{-\text{i}(E_l-E_{l^\prime})t}\ket{\Phi_l}\bra{\Phi_{l^\prime}}\nonumber\\
    &\approx& \sum_{l,l^\prime}c_lc_{l^\prime}^\star\text{e}^{-\text{i}(E_l^{(1)}-E^{(1)}_{l^\prime})t}\ket{\Phi_l^0}\bra{\Phi^0_{l^\prime}}\nonumber\\
    &=&\tilde{\rho},
    \end{eqnarray}
    up to a correction of $\mathcal{O}(1/J_\perp)$ as long as  $t\leq J_\perp$. 

Note that all the density matrices which we described above, have the property that
\begin{equation}
    \rho \approx P_g \rho P_g,
\end{equation}
implying that these density matrices are approximately the same as their projection onto the low energy sector.

In this paper, we focus on a number of \emph{typical} Hermitian operators and entanglement in the 2D model, and ask to what extent their quantitative as well as qualitative behaviour are approximated by the 1D $XXZ$ model.  For this investigation, it is important to work out the effective couplings of $\tilde{H}$ in terms of the original couplings in $H$, which depends explicitly on the boundary conditions of the 2D lattice. In this respect, one may consider four possible cases -- (1) PBC along both rungs and legs, (2) PBC along the rugs and OBC along the legs, (3) OBC along the rungs and PBC along the legs, and (4) OBC along both rungs and legs, all of which we discuss in detail in Sec.~\ref{sec:details}.   Readers interested in the properties of expectation values of relevant operators and entanglement can skip Sec.~\ref{sec:details}, and proceed directly to Secs.~\ref{sec:mapping_observables} and \ref{sec:entanglement}, where these results are discussed. 

\section{Working out the effective Hamiltonian}
\label{sec:details}

In this section, we explicitly work out the low-energy effective Hamiltonian (\ref{eq:1d_general}) for different boundary conditions along the legs and rungs. For brevity, we have assumed $J_\perp=1$ only for this section.

\subsection{Periodic boundary condition along the rungs}
\label{subsec:pbc_along_rungs}

For the case of PBC along the rungs, given by $\vec{\sigma}_{i,L+1}\equiv\vec{\sigma}_{i,1}$, 
we present below analytical forms of the ground states $\ket{\psi^{(i)}_{k}} , k = 0,1$ of the rung Hamiltonian $H_{R_i}$. This in turn helps us to get analytic expressions for the couplings of effective Hamiltonian. 

From the Hamiltonian $H_{R_i}$ given in Eq.~(\ref{eq:system_Hamiltonian_one_rung}) (recall that we have set $J_\perp =1$), it is obvious that for large $h$, the minimum energy state is given by all up spins
\begin{eqnarray}
\ket{\psi_0^{(i)}}=\otimes_{j=1}^L\ket{0_{i,j}}, 
\label{eq:all_up}
\end{eqnarray}
having energy $L\left(\frac{1}{4}-\frac{h}{2}\right)-E_g$. The next higher energy states at large $h$ are given by one spin flip - since this can occur at any of the $L$ rung sites, there are $L$ such states. Since the Hamiltonian $H_{R_i}$ is translationally invariant, we can switch to momentum basis via 
\begin{equation}
    \ket{q}=\frac{1}{\sqrt{L}}\sum_{j=1}^{L}\text{e}^{2iqj\pi/L}\ \sigma^x_{i,j} \ket{\psi_0^{(i)}},
\end{equation}
where $q=0,1,2, \dots ,L-1$. These states are energy eigenstates for all values of $h$, i.e 
\begin{eqnarray}
    H_{R_i}\ket{q}&=&\Big[\cos\Big(\frac{2\pi q}{L}\Big)+L\Big(\frac{1}{4}-\frac{h}{2}\Big)+h-1-E_g\Big]\ket{q}.\nonumber\\  
\end{eqnarray}
For even $L$\footnote{For odd $L$, the minimum energy states in this sector correspond to $(L\pm1)/2$. This leads to a three-fold degeneracy of ground states of the rung at $h=h^\prime$, which is beyond the scope of our consideration in this paper.}, the minimum energy eigenstate corresponds to $q=L/2$, with energy eigenvalue $L\left(\frac{1}{4}-\frac{h}{2}\right)+h-2-E_g$. This state becomes degenerate with the minimum energy state at  
\begin{eqnarray}
h^\prime=2, 
\label{eq:field-strength_for_degeneracy_pbc}
\end{eqnarray}
while the state itself is given by
\begin{eqnarray}
\ket{\psi_1^{(i)}}=\ket{q=L/2}=\frac{1}{\sqrt{L}}\sum_{j=1}^{L}(-1)^{j}\ket{j}. 
\label{eq:all_up_except_one}
\end{eqnarray}
We must comment here that while it is reasonable to expect that the first excited state at large $h$ becomes degenerate with the minimum energy state as we dial down $h$, it is not guaranteed to be. However, our numerical investigation verifies this expectation to be correct.

In order to compute parameters of the effective 1d XXZ $\tilde{J}_{xy}$, $\tilde{J}_{zz}$, $\tilde{h}$ (see Eq.~(\ref{eq:1d_general})), we will match the matrix elements of $H$ and $\tilde H$ in the following states:
\begin{eqnarray}
\ket{\Psi_0}&=&\otimes_{i=1}^N\ket{\psi_0^{(i)}},\nonumber \\
\ket{\Psi_j}&=&\ket{\psi_{1}^{(j)}}\otimes_{\underset{i\neq j}{i=1}}^N\ket{\psi_0^{(i)}},\nonumber\\
\ket{\Psi_{j,j+1}}&=&\ket{\psi_{1}^{(j)}}\otimes \ket{\psi_{1}^{(j+1)}}\otimes_{\underset{i\neq j,j+1}{i=1}}^N\ket{\psi_0^{(i)}}.
\end{eqnarray}
By matching the matrix elements, the expressions for $\tilde{J}_{xy}$, $\tilde{J}_{zz}$, and $\tilde{h}$ can be obtained in terms of the original couplings of the model for PBC along the legs ($\vec{\sigma}_{N+1,j}\equiv\vec{\sigma}_{1,j}$) as  
\begin{eqnarray}
    \tilde{J}_{xy}&=&(J_{||}-J_{d}^{1}-J_{d}^{2})/4, \nonumber \\
    \tilde{J}_{zz}&=&(J_{||}+J_{d}^{1}+J_{d}^{2})/4L, \nonumber \\
    \tilde{h}&=&\left((L-1)(J_{||}+J_{d}^{1}+J_{d}^{2})-L\Delta h\right)/2L. 
\label{eq:effective_coupling_closed-closed}    
\end{eqnarray}
Note that $\tilde{J}_{xy}$, $\tilde{J}_{zz}$, and $\tilde{h}$ are dimensionless numbers, since the $J_\perp^{-1}$ factor, having value $1$, is implicit in all expressions.

Similar mapping and analysis can also be done when the legs obey open boundary condition (OBC). In this case, the effective 1D model is given by 
\begin{eqnarray}
\tilde{H}_{\text{OBC}} &=& \tilde{H}+\tilde{h}^\prime(\tau^z_1+\tau^z_N),
\label{eq:1d_general_obc}
\end{eqnarray}
with $\tilde{H}$ as in Eq.~(\ref{eq:1d_general}) with its effective couplings given in Eq.~(\ref{eq:effective_coupling_closed-closed}), 
the edge-inhomogeneity in the field strength as 
\begin{eqnarray}
    \tilde{h}^\prime&=& -(L-1)(J_{||}+J_{d}^{1}+J_{d}^{2})/4L.
\label{eq:effective_coupling_closed-open_field}     
\end{eqnarray}

Note that for both PBC and OBC along the legs, in the limit $L\rightarrow\infty$, $\tilde{J}_{zz}\rightarrow 0$, implying that the 2D model is mapped to an effective 1D XX model in a transverse magnetic field~\cite{Lieb1961,Barouch1970,Barouch1971,Barouch1971a}, which can be exactly solved via fermionization using the Jordan-Wigner transformation, followed by a Fourier transformation. 

\begin{figure*}
\includegraphics[width=0.8\textwidth]{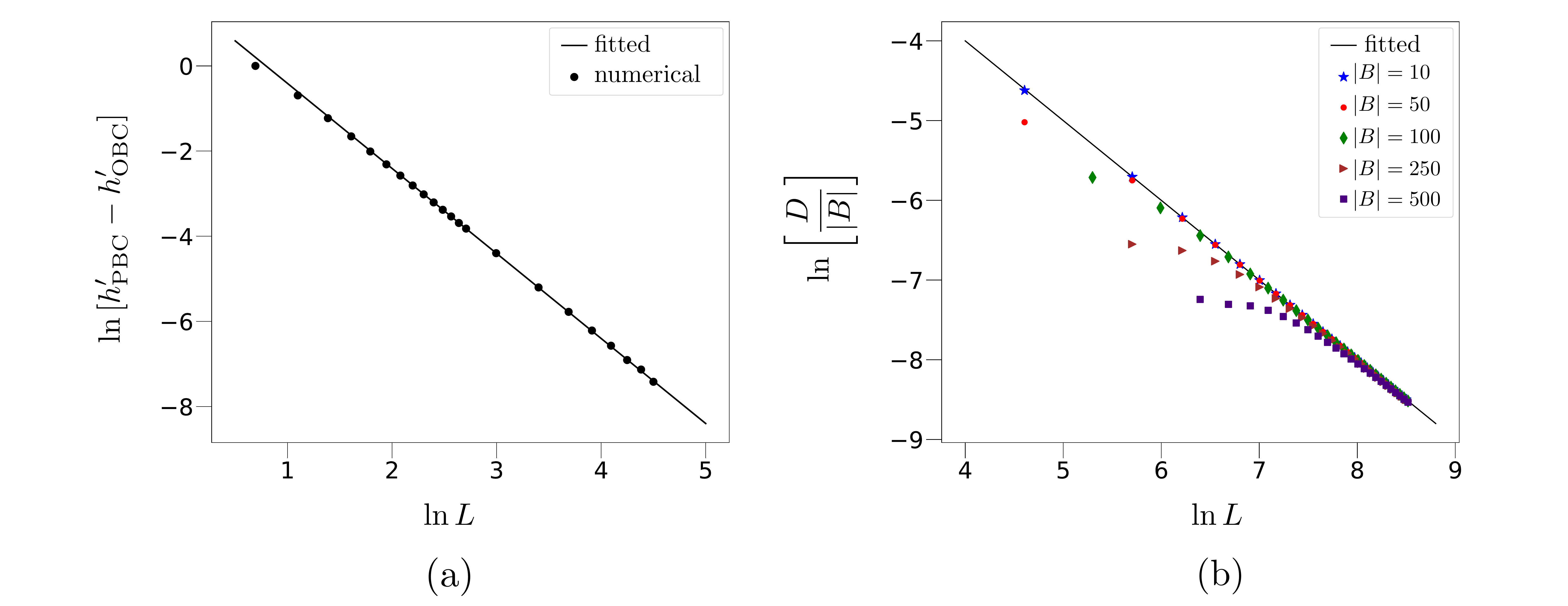}
\caption{(a) Variation of $\ln\left(h^\prime_{\text{PBC}}-h^\prime_{\text{OBC}}\right)$ as a function of $\ln L$. The numerical data is obtained by considering $2\leq L\leq 10^2$, and is fitted to Eq.~(\ref{eq:h_fit}) with $a=4.88251$, and the power of $L$ obtained as $\sim -1.99684$ from the fitted curve. (b) Variation of $\ln\left[D/|B|\right]$ as a function of $\ln L$ with different values of $|B|$. The numerical data is obtained by considering $10^2\leq L\leq 5\times 10^3$, and is fitted to Eq.~(\ref{eq:bulk_fit}), where the power of $L$ is obtained as $\sim-1.004$ from the fitted curve. All quantities plotted in all figures are dimensionless.}
\label{fig:fig2_scaling}
\end{figure*}

\subsection{Open boundary condition along the rungs}
\label{subsec:obc_along_rungs}

Open boundary condition (OBC) along the rungs represents a geometry of the lattice that is different from the one discussed in Sec.~\ref{subsec:pbc_along_rungs}. Exact calculation of the degenerate ground states and the strength of the magnetic field at which the degeneracy occurs is difficult for arbitrary $L$. However, our numerical analysis suggests that for open (periodic) boundary condition along the rungs (legs), and for arbitrary $L$, $\ket{\psi_0^{(i)}}$ is still given by Eq.~(\ref{eq:all_up}), while  $\ket{\psi_1^{(i)}}$ is found in the $\langle Z_i\rangle=L-2$ sector, and has the form 
\begin{eqnarray}
\ket{\psi_1^{(i)}} &=& \sum_{j=1}^L a_j \ket{j}, 
\label{eq:all_up_except_one_obc}
\end{eqnarray}
with $a_j\in\mathbb{R}$, and $\sum_{j=1}^La_j^2=1$. Note further that due to $Z_2$ symmetry, $a_1=\pm a_L$. Using Eqs.~(\ref{eq:all_up}) and (\ref{eq:all_up_except_one_obc}), the effective couplings in Eq.~(\ref{eq:1d_general}) can be determined as 
\begin{eqnarray}
\tilde{J}_{xy} &=&[J_{||}+A_0(J_{d}^{1}+J_{d}^{2})]/4,\nonumber\\
\tilde{J}_{zz} &=&[(4-L+A_1)J_{||}\nonumber\\&&+(L-1+A_3-2A_2)(J_d^1+J_d^2)]/16,\nonumber \\
\tilde{h} &=&[(L-A_1)J_{||}+(L-1-A_3)(J_d^{1}+J_{d}^{2})-2\Delta h]/8,\nonumber\\
\label{eq:effective_coupling_open_closed}
\end{eqnarray}
with 
\begin{eqnarray}
\label{eq:A0}
A_0&=& \sum_{j=1}^{L-1}a_ja_{j+1},\\
\label{eq:A1}
A_1&=&L\sum_{j=1}^L a_j^4+(2L-8)\sum_{j=1}^L\sum_{k<j}a_j^2a_k^2,\\
\label{eq:A2}
A_2&=&L-3+2a_1^2,\\ 
\label{eq:A3}
A_3 &=&(L-1)\Big\{\sum_{j=1}^{L-1}a_{j}^{2}a_{j+1}^{2}+a_{L}^{2}a_{1}^{2}\Big\}\nonumber\\ &&
+(L-3)\Big\{\sum_{j=1}^{L-1}a_{j}^{2}a_{1}^{2}+\sum_{j=2}^{L}a_{L}^{2}a_{j}^{2}\Big\} \nonumber \\
&&
+(L-5)\Big\{\sum_{j=1}^{L-1}\sum_{k=j+2}^{L}a_{j}^{2}a_{k}^{2} +\sum_{j=2}^{L}\sum_{k=j}^{L-1}a_{k}^{2}a_{j}^{2}\Big\}.
\end{eqnarray}
The calculation is involving, and we have demonstrated the important steps in Appendix~\ref{app:obc_along_rungs_coefficients}.

Similar calculation can also be performed for the case of OBC along the legs, where the 1D LEH picks up an inhomogeneity of the effective magnetic field strength $\tilde{h}$ at the edge spins,  so that the effective Hamiltonian now reads (\ref{eq:1d_general_obc}), where the effective couplings are given in Eq.~(\ref{eq:effective_coupling_open_closed}),  
the edge-inhomogeneity in the field is 
\begin{eqnarray}
\tilde{h}^{\prime} &=&-[(L-1-A_3)(J_d^{1}+J_{d}^{2})+(L-A_1)J_{||}]/16.
\label{eq:effective_couplings_open_open_field}
\end{eqnarray}

\subsubsection{Large \texorpdfstring{$L$}{L} limit}

At $L\rightarrow\infty$, the field-strength $h^\prime$ leading to doubly degenerate ground states of the rung approach the same in the case of PBC along the rungs. To demonstrate this numerically, we show that $h^\prime_{\text{OBC}}$ asymptotically approaches $h^\prime_{\text{PBC}}=2$ (see Eq.~(\ref{eq:field-strength_for_degeneracy_pbc})) as 
\begin{eqnarray}
h^\prime_{\text{OBC}} &=& h^\prime_{\text{PBC}}-\left(\frac{a}{L}\right)^2+\cdots,
\label{eq:h_fit}
\end{eqnarray}
where $a$ is a constant, and  we ignore third and higher order terms in $1/L$ (see Fig.~\ref{fig:fig2_scaling}(a) for a demonstration).  

In order to use the ground state $\ket{\psi_1^{(i)}}$ of the $i$th rung with PBC as an approximation of the same with OBC along the rungs in the $L\rightarrow\infty$ limit, it is important to investigate up to what extent the former mimics the latter. In this paper, we are interested in the \emph{bulk} of the system located at the middle of the lattice. This naturally includes the bulk $B$ of the rungs at the middle of the lattice, the size $|B|$ of which is determined by the lattice sites included in it. The reduced state $\rho_B$ of the bulk of a rung can be determined by tracing out all spins from $\ket{\psi_1^{(i)}}$ except the ones in the bulk. To quantitatively estimate how $\rho_B^{\text{OBC}}$ corresponding to OBC along rungs approaches $\rho_{B}^{\text{PBC}}$ when the rungs form closed chains, we focus on the trace distance~\cite{wilde_book} (see Appendix~\ref{app:distance} for the definition), $D$, between $\rho_B^{\text{PBC}}$ and $\rho_B^{\text{OBC}}$. Our numerical results indicate that $D$ varies with $L$ as 
\begin{eqnarray}
D &\sim& \frac{|B|}{L}, 
\label{eq:bulk_fit}
\end{eqnarray}
and tends to $0$ for $L\rightarrow\infty$. Here, second and higher powers of $1/L$ are ignored. See Fig.~\ref{fig:fig2_scaling}(b) for a demonstration. This result indicate that for $L\rightarrow\infty$, a rung with OBC can be reliably mimicked by a rung with PBC as long as the investigation is confined in the bulk of the chain.

\begin{table}
\begin{tabular}{|c|c|}
\hline 
$A$ & $\tilde{A}$ \\
\hline \hline 
$\sigma_{i,j}^\alpha$ & $\frac{(-1)^j}{\sqrt{L}}\tau_i^\alpha$, for $\alpha=x,y$\\
\hline
$\sigma_{i,j}^z$ & $\frac{L-1}{L}I_i+\frac{1}{L}\tau^z_i$\\
\hline 
$\sum_{j=1}^L \sigma^{z}_{i,j}$ & $(L-1)I_i+\tau^z_i$\\
\hline
$\sum_{j=1}^L \sigma^{x}_{i,j}$ & $0$\\
\hline 
$\sum_{j=1}^L \sigma^{y}_{i,j}$ & $0$ \\
\hline 
$\otimes_{j=1}^L\sigma^z_{i,j}$ & $\tau^z_i$ \\
\hline 
$\otimes_{j=1}^L\sigma^x_{i,j}$ & \begin{tabular}{c} $\frac{1}{2}(\tau_i^z-I_i)$ for $L=2$ \\  0 for $L>2$ \end{tabular} \\
\hline 
$\otimes_{j=1}^L\sigma^y_{i,j}$ & \begin{tabular}{c} $\frac{1}{2}(\tau_i^z-I_i)$ for $L=2$ \\  0 for $L>2$ \end{tabular} \\
\hline 
$\sigma^x_{i,1}\otimes\sigma^z_{i,2}$ & $-\frac{1}{\sqrt{2}}\tau^x_i$ \\
\hline 
$\sigma^z_{i,1}\otimes\sigma^x_{i,2}$ & $\frac{1}{\sqrt{2}}\tau^x_i$\\
\hline 
$\sigma^y_{i,1}\otimes\sigma^z_{i,2}$ & $-\frac{1}{\sqrt{2}}\tau^y_i$ \\
\hline 
$\sigma^z_{i,1}\otimes\sigma^y_{i,2}$ & $\frac{1}{\sqrt{2}}\tau^y_i$\\
\hline 
$\sigma^x_{i,1}\otimes\sigma^y_{i,2}$ & $0$ \\
\hline 
$\sigma^y_{i,1}\otimes\sigma^x_{i,2}$ & $0$\\
\hline 
$\sigma^\alpha_{i,j}\otimes\sigma^\alpha_{i+r,j^\prime}$ & $\frac{(-1)^{j+j^\prime}}{L}\tau^\alpha_i\otimes\tau^\alpha_{i+r}$ for $\alpha=x,y$\\
\hline 
$\sigma^z_{i,j}\otimes\sigma^z_{i+r,j^\prime}$ & $\frac{(L-1)^2}{L^2} I+\frac{1}{L^2}\tau^z_i\tau^z_{i+r}+\frac{L-1}{L^2}(\tau^z_i+\tau^z_{i+r})$ \\
\hline 
$\sigma^x_{i,j}\otimes\sigma^z_{i+r,j^\prime}$ & $\frac{(-1)^j}{\sqrt{L}}\left[\frac{L-1}{L}\tau^x_i+\frac{1}{L}\tau^x_i\tau^z_{i+r}\right]$ \\
\hline 
$\sigma^z_{i,j}\otimes\sigma^x_{i+r,j^\prime}$ & $\frac{(-1)^{j^\prime}}{\sqrt{L}}\left[\frac{L-1}{L}\tau^x_{i+r}+\frac{1}{L}\tau^z_i\tau^x_{i+r}\right]$ \\
\hline
$\sigma^y_{i,j}\otimes\sigma^z_{i+r,j^\prime}$ & $\frac{(-1)^j}{\sqrt{L}}\left[\frac{L-1}{L}\tau^y_i+\frac{1}{L}\tau^y_i\tau^z_{i+r}\right]$ \\
\hline
$\sigma^z_{i,j}\otimes\sigma^z_{i+r,j^\prime}$ & $\frac{(-1)^{j^\prime}}{\sqrt{L}}\left[\frac{L-1}{L}\tau^y_{i+r}+\frac{1}{L}\tau^z_i\tau^y_{i+r}\right]$ \\
\hline
$\sigma^x_{i,j}\otimes\sigma^y_{i+r,j^\prime}$ & $0$ \\
\hline
$\sigma^y_{i,j}\otimes\sigma^x_{i+r,j^\prime}$ & $0$ \\
\hline
\end{tabular}
\caption{A list of typical observables on the ladder and the corresponding low-energy operators on the 1D effective $XXZ$ model, where PBC along both rungs and legs have been used for calculation. A subset of these operators for $L=2$ can be found in~\cite{Mila1998}.}
\label{operators}
\end{table}

\subsubsection{Mapping for odd \texorpdfstring{$L$}{L} with vanishing magnetic field}

We now discuss a special case where $L$ is odd, and no magnetic field is present in the system, i.e., $h=0$. In this situation, each rung $i$ of the system has a doubly degenerate ground state having the form 
\begin{eqnarray}
\ket{\psi_0^{(i)}} &=& \sum_{j}a_j\mathcal{P}_j\left(\ket{0}^{\otimes (L+1)/2}\ket{1}^{\otimes (L-1)/2}\right), 
\label{eq:0state0}
\end{eqnarray}
with $\sum_j a_j^2=1$, 
and 
\begin{eqnarray}
\ket{\psi_1^{(i)}} &=& \otimes_{j=1}^L\sigma^x_j\ket{\psi_0^{(i)}}.
\label{eq:0state1}
\end{eqnarray}
Here, $\{\mathcal{P}_j\}$ is the set of all possible permutations of $L^\prime$ ($L-L^\prime$) spins at ground (excited) state, with $L^\prime=(L-1)/2$ $\left(L^\prime=(L+1)/2\right)$ for $\ket{\psi_0^{(i)}}$ $(\ket{\psi_1^{(i)}})$.  \textbf{Proposition 1} (see Sec.~\ref{subsec:leh}) is valid in this case also, and calculation (see Appendix~\ref{app:zero_field}) leads to an effective 1D model given by the Hamiltonian
\begin{eqnarray}
\tilde{H} &=& \sum_{i=1}^{N}\left[\tilde{J}_{xy}\left(\tau_i^x\tau_{i+1}^x+\tau_i^y\tau_{i+1}^y\right)+\tilde{J}_{zz}\tau_i^z\tau_{i+1}^z\right]. 
\label{eq:1d_leh_zero_field}
\end{eqnarray}
We demonstrate the important steps of the calculation in Appendix~\ref{app:zero_field}, where we set $J_d^1=J_d^2=0$ to keep the calculation uncluttered. However, similar results can be obtained also for non-zero perturbations $J_d^1,J_d^2$, where the overall form of Eq.~(\ref{eq:1d_leh_zero_field}) remains unchanged.

\begin{figure}
\includegraphics[width=0.8\linewidth]{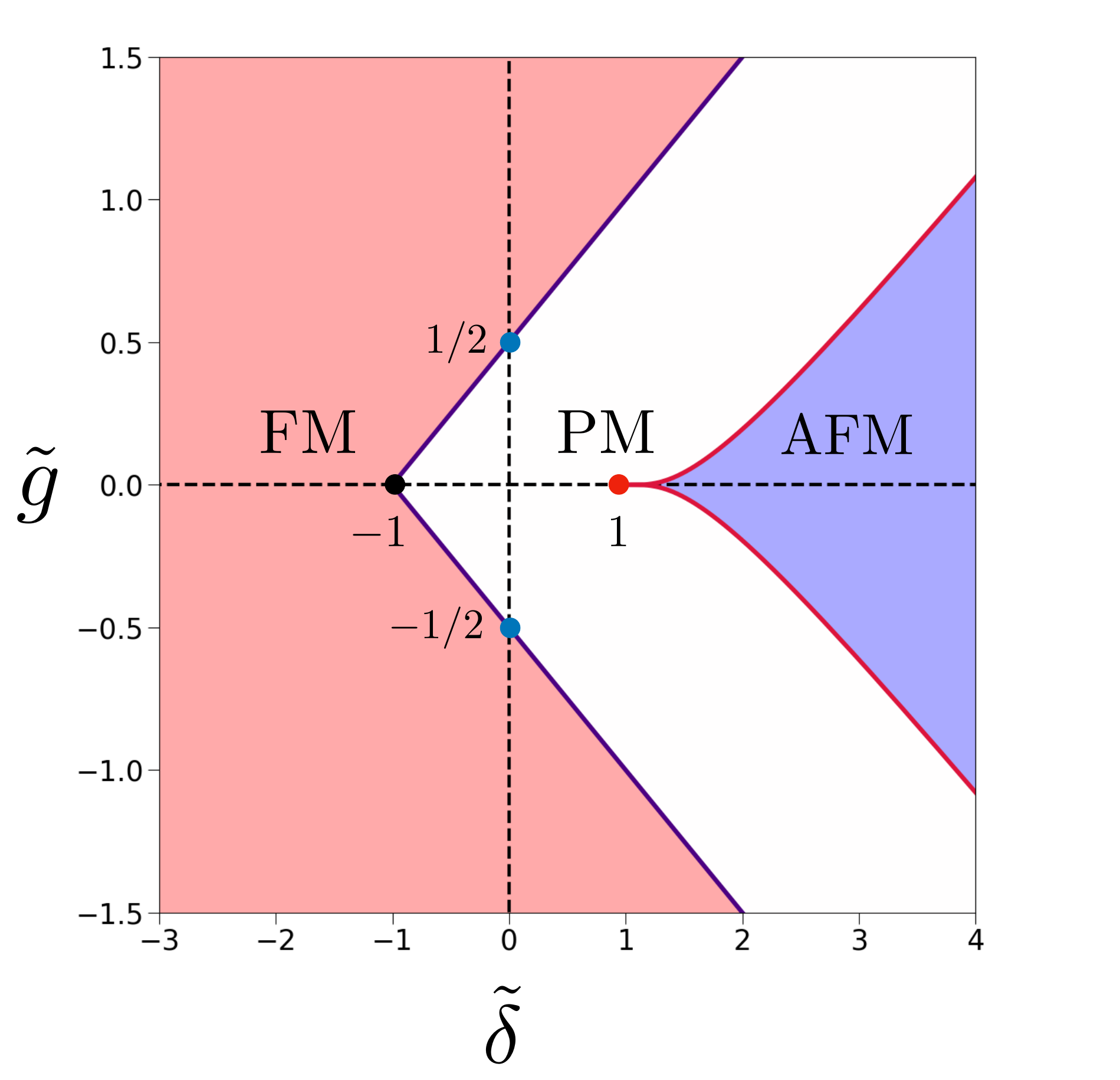}
\caption{Phase diagram of the 1D $XXZ$ model in a magnetic field on the plane of $(\tilde{\delta},\tilde{g})$~\cite{Mikeska2004,Franchini2017}, marking the antiferromagnetic (AFM), ferromagnetic (FM), and paramagnetic (PM) phases and the corresponding phase boundaries, where $\tilde{\delta}=\tilde{J}_{zz}/\tilde{J}_{xy}$ quantifies the $z$-anisotropy, and $\tilde{g}=\tilde{h}/\tilde{J}_{xy}$.}
\label{fig:phase_diagram}
\end{figure}

\section{Mapping observables}
\label{sec:mapping_observables}

It has been seen in Sec.~\ref{sec:methodology} that specific states in the low-energy sector of the 2D model (\ref{eq:system_Hamiltonian}) is approximated, up to corrections of $\mathcal{O}(1/J_\perp)$, by the spectrum of the 1D effective $XXZ$ model. A natural question that arises at this point is \emph{whether the low energy content of an observable (i.e., matrix elements of Hermitian operators in the low energy subspace) are captured by an observable defined entirely on the Hilbert space of the 1D effective $XXZ$ model?}. We answer this question affirmatively below. 

To put the question in a more concrete mathematical footing, we focus on the matrix element $\langle\Phi_l|A|\Phi_{l^\prime}\rangle$ of a Hermitian operator $A$ on the space of the non-degenerate low energy states of the 2D model, 
and ask whether there exists a Hermitian operator $\tilde{A}$ on the space of the 1D $XXZ$ model such that
\begin{eqnarray}
\langle\Phi_l|A|\Phi_{l^\prime}\rangle\approx\langle\Phi_l^0|\tilde{A}|\Phi_{l^\prime}^0\rangle,
\label{eq:operator_approximation}
\end{eqnarray}
Towards answering this question, we note that an alternative definition for $P_g$ would be $P_g=\sum_{l}\ket{\Phi_l^0}\bra{\Phi_l^0}$, which is better suited for subsequent discussion. Equipped with this, we propose the following.  

\noindent\textbf{Proposition 2.} \emph{Eq.~(\ref{eq:operator_approximation}) holds if $\tilde{A}=P_gAP_g$}.

\begin{proof} 
To prove this, let us write
\begin{eqnarray}
\langle\Phi_l|A|\Phi_{l^\prime}\rangle &=& \langle\Phi_l|P_g A P_g|\Phi_{l^\prime}\rangle + \langle\Phi_l|P_e A P_g|\Phi_{l^\prime}\rangle \nonumber\\ &+& \langle\Phi_l|P_g A P_e|\Phi_{l^\prime}\rangle+ \langle\Phi_l|P_e A P_e|\Phi_{l^\prime}\rangle,
\label{eq:operator_mapping}
\end{eqnarray} 
and note that for the low-energy manifold of $H$, 
\begin{eqnarray}
P_g\ket{\Phi_{l^\prime}}=\sum_l\ket{\Phi_l^0}\langle\Phi_l^0|\Phi_{l^\prime}\rangle\approx \ket{ \Phi_{l^\prime}^0 },
\end{eqnarray}
since for $l\neq l^\prime$, $\langle\Phi_l^0|\Phi_{l^\prime}\rangle\rightarrow 0$ due to Eq.~(\ref{eq:State_perturbative_corrections}) and  orthonormality of $\ket{\Phi_l^0}$. Also note that $P_e\ket{\Phi_{l^\prime}} \approx P_e\ket{\Phi^0_{l^\prime}} = 0 $. Therefore, 
\begin{eqnarray}
\langle\Phi_l|A|\Phi_{l^\prime}\rangle\approx \langle\Phi_l|P_gAP_g|\Phi_{l^\prime}\rangle= \langle\Phi_l^0|\tilde{A}|\Phi_{l^\prime}^0\rangle,
\end{eqnarray}
where $\tilde{A}=P_gAP_g$. Hence proved. 
\end{proof}
\noindent We point out here that although the proof is presented for non-degenerate eigenstates, it can be straightforwardly extended to the states constructed out of the non-degenerate eigenstates discussed in Sec.~\ref{subsec:merit_of_mapping}.

We now focus on a number of relevant operators $A$ on the space of the 2D model, and determine the corresponding operators $\tilde{A}$ for the 1D effective model. We assume PBC along both rungs and legs for demonstration. However, the prescription can also be applied to other boundary conditions along legs and rungs. We consider mainly two types of operators -- (a) operators $A$ having support only on one rung, say, $i$, where the corresponding operator $\tilde{A}_i^\alpha=P_gA_iP_g$ is a $2\times 2$ matrix in the $\{\ket{\mathbf{0}},\ket{\mathbf{1}}\}$ basis, and has support on only the $i$th lattice site of the 1D $XXZ$ model, and (2) operators of the form $A_{i,i^\prime} = A_{i} \otimes A_{i^\prime}$ having support on two different rungs, $i$, and $i^\prime$, where the mapped operator $\tilde{A}_{i,i^\prime}$ is a $4\times 4$ Hermitian matrix in the basis $\{\ket{\mathbf{00}},\ket{\mathbf{01}},\ket{\mathbf{10}},\ket{\mathbf{11}}\}$ on the spins $(i,i^\prime)$ in the 1D $XXZ$ spin-chain. To determine the low-energy component of operators of the latter type, we note the following.

\begin{figure*}
    \includegraphics[width=0.8\textwidth]{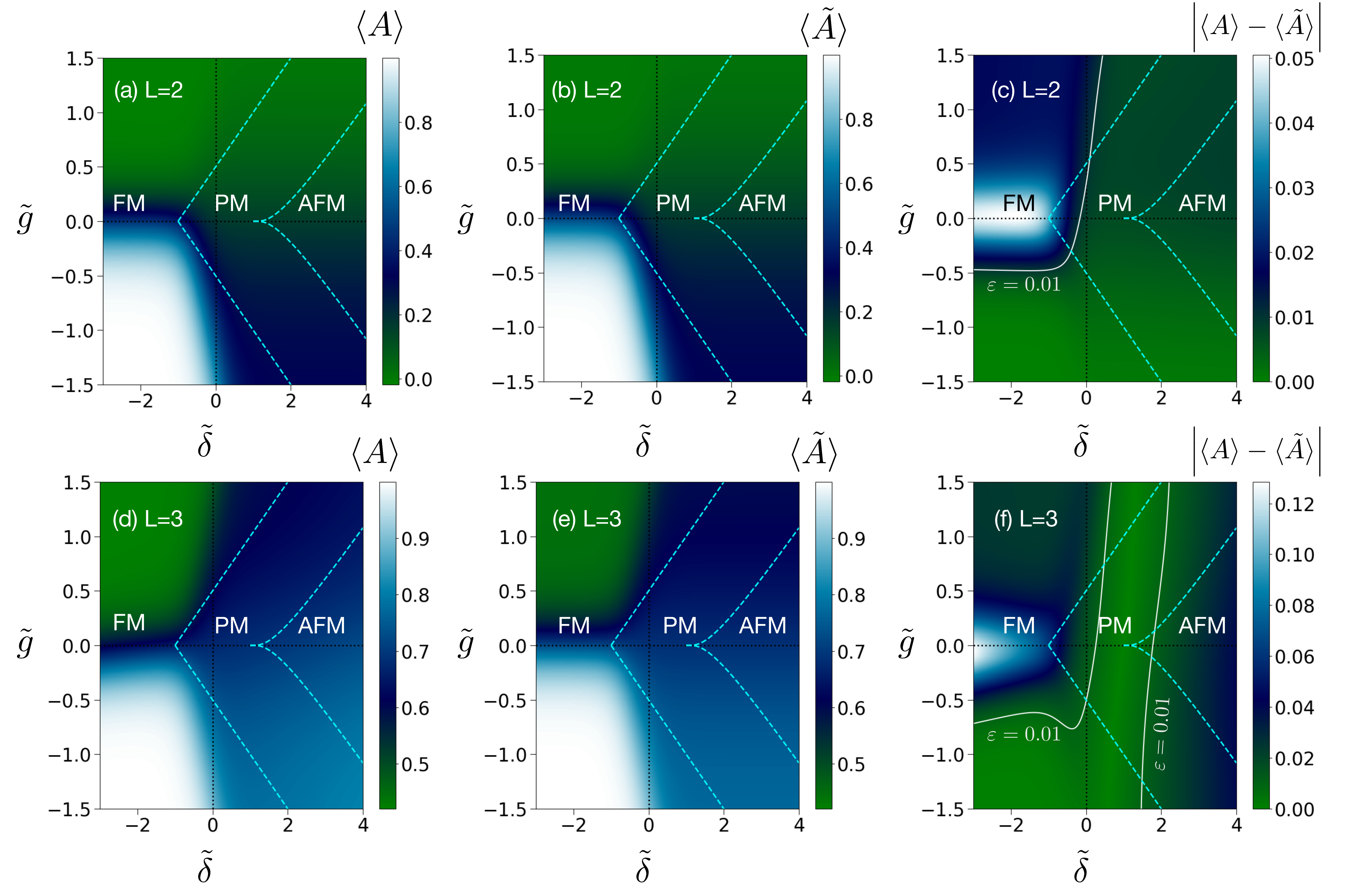}
    \caption{Variations of $\langle A\rangle$, $\langle \tilde{A}\rangle$, and $\varepsilon=|\langle A\rangle-\langle \tilde{A}\rangle|$ as functions of $\tilde{\delta}$ and $\tilde{g}$, where $A=\sigma^z_{i,1}\otimes\sigma^z_{i+1,1}$, $\tilde{A}=(I+\tau^z_i+\tau^z_{i+1}+\tau^z_i\otimes\tau^z_{i+1})/4$ for $L=2$, and $\tilde{A}=[25I+5(\tau^z_i+\tau^z_{i+1})+\tau^z_i\otimes\tau^z_{i+1}]/36$ for $L=3$. All quantities plotted are dimensionless.}
    \label{fig:observables_zz_nn}
\end{figure*}

\noindent\textbf{Proposition 3.} \emph{For an operator on the space of the 2D model having the form $A_{i,i^\prime}=A_{i}\otimes A_{i^\prime}$ with $A_i$ and $A_{i^\prime}$ having support only on rungs $i$ and $i^\prime$ respectively,  $\tilde{A}_{i,i^\prime}=P_gA_{i,i^\prime}P_g=\tilde{A}_i\otimes\tilde{A}_{i^\prime}$, where $\tilde{A}_{i}=P_gA_iP_g$, and $\tilde{A}_{i^\prime}=P_gA_{i^\prime}P_g$}.

\begin{proof}
To prove this, we write $P_g=\otimes_{i=1}^NP_g^{(i)}$, with $P_g^{(i)}=\sum_{k_i=0}^1\ket{\psi_{k_i}^{(i)}}\bra{\psi_{k_i}^{(i)}}$, where we use notations introduced in Eq.~(\ref{eq:ground_state_manifold}). This leads to 
\begin{eqnarray}\label{eq:Product of operators effective description}
\tilde{A}_{i,i^\prime}&=&P_g A_{i,i^\prime}P_g\nonumber\\ 
&=&[P_g^{(i)}\otimes P_g^{(i^\prime)}](A_i\otimes A_{i^\prime})[P_g^{(i)}\otimes P_g^{(i^\prime)}]\nonumber\\ 
&=&[P_g^{(i)}A_iP_g^{(i)}]\otimes[P_g^{(i^\prime)}A_{i^\prime}P_g^{(i^\prime)}]\nonumber\\
&=&\tilde{A}_i\otimes\tilde{A}_{i^\prime}.
\end{eqnarray}
Here we have been explicit only about projectors in the rungs $i$ and $i'$. Hence the proof.
\end{proof}
\noindent Note that this proposition can be extended to any operator on the space of the 2D model, since it can always be written as a linear combination of tensor products of operators localized on rungs.

We now specifically consider four constructions of $A$, namely, $\sigma^\alpha_{i,j}$, $\sum_{j=1}^L\sigma^{\alpha}_{i,j}$, $\otimes_{j=1}^L\sigma^{\alpha_j}_{i,j}$\footnote{One might naively guess that if $A_i,B_i$ maps to $\tilde A_i,\tilde B_i$ in $XXZ$ model, then for $C_i = A_i B_i$ the corresponding $XXZ$ operator is $\tilde C_i = \tilde A_i \tilde B_i$. However this is not true - see Eq.~(\ref{eq:Product of operators effective description}) for the caveat.}, and $\sigma^{\alpha_i}_{i,j}\otimes\sigma^{\alpha_{i+r}}_{i+r,j^\prime}$, where $\alpha,\alpha_i\in\{x,y,z\}$. The expressions for the corresponding low-energy components $\tilde{A}$ on the 1D $XXZ$ model are given in Table~\ref{operators}. For ease of reference, we call the operators $A$ having vanishing low-energy component (i.e., $\tilde{A}=0$) as \emph{high-energy operators}. Note here that the low-energy component of $\sigma^\alpha_{i,j}$, $\alpha=x,y,z$, is consistent with the operator identification of $J_i^z$ in terms of $\tau_i^z$, as given in Eq.~(\ref{eq:Action of J_z}) and Eq.~(\ref{eq:epauliz}). 

To systematically test the performance of the first-order approximation for the operators listed in Table~\ref{operators}, 
we consider the situation of PBC along the legs\footnote{OBC along the legs can also be considered in similar fashion.}. Note that the 1D effective model can be described by two independent dimensionless parameters, $\tilde{g}=\tilde{h}/\tilde{J}_{xy}$, and $\tilde{\delta}=\tilde{J}_{zz}/\tilde{J}_{xy}$, where  the dimensionless parameter $\tilde{J}_{xy}\neq 0$, and the physics of the model is expected to be invariant with a change in the value of $\tilde{J}_{xy}$. Using $\tilde{g}$ and $\tilde{\delta}$, the phase diagram of the 1D $XXZ$ model is given by Fig.~\ref{fig:phase_diagram}, where the ferromagnetic (FM), antiferromagnetic (AFM), and the paramagnetic (PM) phases and the corresponding phase boundaries are marked. We use this phase diagram for reference during discussion of results in  subsequent sections. The effective coupling constants are functions of $J_{||}/J_\perp$, $J_{d}^{1,2}/J_\perp$, and $\Delta h/\perp$ (see Eqs.~(\ref{eq:effective_coupling_closed-closed}) and (\ref{eq:effective_coupling_open_closed})).  Let us define $J_d^{\text{sum}}=|(J_d^1+J_d^2)/J_\perp|$, $J_d^{\text{diff}}=|(J_d^1-J_d^2)/J_\perp|$, and observe that $\tilde{J}_{xy}$, $\tilde{J}_{zz}$, and $\tilde{h}$ depend only on $J_d^{\text{sum}}$, and are independent of $J_d^{\text{diff}}$. Solving for $J_{||}/J_\perp$, $J_d^{\text{sum}}/J_\perp$, and $\Delta h/J_\perp$ from Eq.~(\ref{eq:effective_coupling_closed-closed}), one can write 
\begin{eqnarray}
\frac{J_{||}}{J_\perp} &=& 2\tilde{J}_{xy}(L\tilde{\delta}+1),\;
\frac{J_d^{\text{sum}}}{J_\perp} = 2\tilde{J}_{xy}(L\tilde{\delta}-1),, \nonumber\\
\frac{\Delta h}{J_\perp} &=& 2\tilde{J}_{xy}[2(L-1)\tilde{\delta}-\tilde{g}],
\end{eqnarray}
for PBC along the rungs. In the case of OBC along the rungs, from Eq.~(\ref{eq:effective_coupling_open_closed}), one obtains 
\begin{eqnarray}
\frac{J_{||}}{J_\perp} &=&\frac{4\tilde{J}_{xy}(4A_0\tilde{\delta}+C_1)}{C_1+A_0C_2},\;
\frac{J_d^{\text{sum}}}{J_\perp} = \frac{4\tilde{J}_{xy}(C_2-4\tilde{\delta})}{C_1+A_0C_2}, \nonumber\\
\frac{\Delta h}{J_\perp} &=& 2\tilde{J}_{xy}\Big[\frac{(4-C_2)(4A_0\tilde{\delta}+C_1)}{C_1+A_0C_2}\nonumber\\ &&
+\frac{2(A_2-A_3)(C_2-4\tilde{\delta})}{C_1+A_0C_2}-4\tilde{g}\Big], 
\end{eqnarray}
where $C_1=2A_2-A_3-L+1$, $C_2=4-L+A_1$, and $A_0$, $A_1$, $A_2$, and $A_3$ are given in Eqs.~(\ref{eq:A0})-(\ref{eq:A3}). In the following, we consider expectation values of specific Hermitian operators, both in the 2D model as well as in the effective 1D model, as functions of $\tilde{g}$ and $\tilde{\delta}$, under the constraints that the corresponding $\left(J_{||}/J_\perp,J_d^{\text{sum}}/J_\perp,J_d^{\text{diff}}/J_\perp,\Delta h/J_\perp\right)$ are small. Note that irrespective of the value of $\tilde{\delta}$ and $\tilde{g}$, this can be ensured by fixing $\tilde{J}_{xy}$ to a small value\footnote{This is the only constraint on $\tilde{J}_{xy}$, which, in turn, implies that $J_{||}/J_\perp$ and $J_d^{\text{sum}}/J_\perp$ have to be comparable.}.  We fix $\tilde{J}_{xy}=10^{-2}$ and $J_{d}^{\text{diff}}=0$ for the range $-3\leq\tilde{\delta}\leq 4$ and $-1.5\leq\tilde{g}\leq 1.5$ considered in this work.    

For demonstration, we choose $A=\sigma^z_{i,1}\otimes\sigma^{z}_{i+1,1}$ on a $3\times 2$ and $3\times 3$ lattice, for which the low-energy component $\tilde{A}$ is given in Table~\ref{operators}. We compute $\langle A\rangle=\text{Tr}(\rho A)$ and $\langle\tilde{A}\rangle=\text{Tr}(\tilde{\rho}\tilde{A})$, where $\rho$ and $\tilde{\rho}$ are considered to be a thermal state (see Eq.~(\ref{eq:thermal_state_H})). To ensure that the state lies in the low-energy sector of the 2D model, we consider $\beta J_\perp=10^2$, where the strength of the rung-coupling is taken to be  $J_\perp=10^2$.  Fig.~\ref{fig:observables_zz_nn} depicts the variation of $\langle A\rangle$, $\langle\tilde{A}\rangle$, and $\varepsilon=|\langle A\rangle-\langle\tilde{A}\rangle|$ as functions of $\tilde{\delta}$ and $\tilde{g}$, where OBC is assumed along the rungs of the lattices. In order to quantitatively estimate the performance of the 1D effective $XXZ$ model to substitute the values of $\langle A\rangle$, on the terrain of $\varepsilon$, we mark the regions in which $\varepsilon\le J_\perp^{-1}=10^{-2}$. Note that in the case of $L=2$, major portions of the considered $(\tilde{\delta},\tilde{g})$ plane is included in this region, with relatively higher values of $\varepsilon$ occurring mostly in the FM phase with $\tilde{g}\geq 0$. In contrast, for $L=3$, higher values of $\varepsilon$ appear in the PM and the AFM phases also. These observations remain qualitatively unchanged for observables defined on a single rung instead of two different rungs. For a discussion and demonstration with $A=\sigma^z_{1,1}\otimes\sigma^z_{i,2}$ and $A=\sigma^z_{i,1}$, see Appendix~\ref{app:observables}. We want to emphasise here that that qualitative variations of $\langle A\rangle$ in the 2D model is mimicked satisfactorily by the same for $\langle\tilde{A}\rangle$ in the effective 1D model, in the entire range of $(\tilde{\delta},\tilde{g})$ considered in this paper, even beyond the realm of perturbation theory.

\begin{figure*}
    \includegraphics[width=0.8\textwidth]{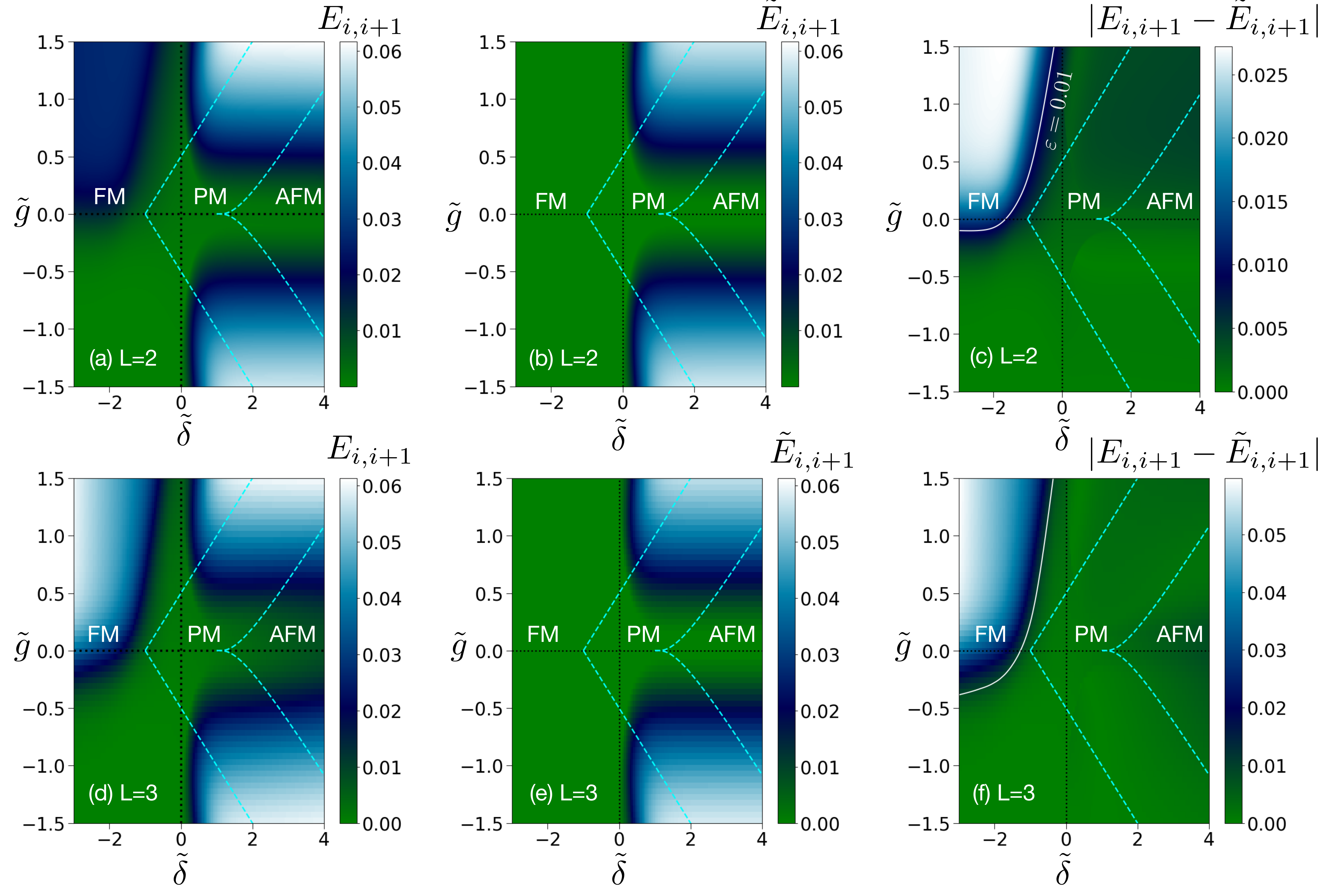}
    \caption{Variations of $E_{i,i+1}$, $ \tilde{E}_{i,i+1}$, and $\varepsilon=|E_{i,i+1}-\tilde{E}_{i,i+1}|$ as functions of $\tilde{\delta}$ and $\tilde{g}$. The top horizontal panel is for a $3\times 2$ lattice, while figures in the bottom horizontal panel are for a $3\times 3$ lattice. All quantities plotted are dimensionless.}
    \label{fig:entanglement_nn}
\end{figure*}

\section{Estimating Entanglement}
\label{sec:entanglement}

Now that we have matched the spectrum and observables from the 2D model to the 1D $XXZ$ model, in this Section, we focus on quantum correlation measures belonging to the entanglement-separability paradigm~\cite{horodecki2009,guhne2009} that can be computed in the system under consideration. 

\subsection{Entanglement from reduced density matrices}
\label{subsec:reduced_density_matrix}

While the distribution of entanglement in a quantum spin model can be varied, such as bipartite and multipartite entanglement over different subsystems, it often adapts a partial trace-based methodology for computation~\cite{horodecki2009}. For example, the bipartite entanglement between two subsystems of the spin model in a pure state, is inferred from the reduced density matrix of one of the subsystems. Therefore, the 
first natural question to address here is if there exists a relation between the reduced density matrices of subsystems of the 2D model, and the reduced density matrices of the corresponding subsystems in the effective 1D $XXZ$ model. In fact, it is natural to expect that a single-rung reduced density matrix in the 2D model will match to a single spin reduced density matrix in the 1D $XXZ$ model. We will now show that this is indeed true. 

Let us begin with a density matrix of the 2D system, which has a description in the 1D effective $XXZ$ model, i.e $\rho\approx P_g\rho P_g =\tilde{\rho}$. For this, we propose the following. 

\noindent$\blacksquare$ \textbf{Proposition 4.} \emph{The reduced density matrix $\rho^{(i)}$ obtained via tracing out the rung $i$ from $\rho$ and the reduced density matrix $\tilde \rho^{(i)}$ obtained from $\tilde{\rho}$ by tracing out the spin $i$ obeys $\rho^{(i)}\approx \tilde{\rho}^{(i)}$.}

\begin{proof} 
Note that exploiting the basis-independence of partial trace, and using notations introduced in Eq.~(\ref{eq:ground_state_manifold}) in Sec.~\ref{subsec:leh},  
\begin{equation}\label{eq:Reduced density  matrix on ladder}
    \rho^{(i)} = \mbox{Tr}_i \rho = \sum_{k=0}^{2^L-1} \langle \psi^{(i)}_k | \rho \ket{ \psi^{(i)}_k }
\end{equation}
whereas 
\begin{equation}\label{eq:Reduced density  matrix on XXZ}
    \tilde \rho^{(i)} = \mbox{Tr}_i \tilde \rho = \sum_{k=0}^{1} \langle \psi^{(i)}_k | \tilde \rho \ket{ \psi^{(i)}_k }
\end{equation}
To prove that $ \rho^{(i)} \approx \tilde \rho^{(i)}$, we consider 
\begin{eqnarray}
    \langle \psi^{(i)}_k | \rho \ket{ \psi^{(i)}_k } \approx \langle \psi^{(i)}_k | \tilde \rho \ket{ \psi^{(i)}_k } = \langle \psi^{(i)}_k | P_g  \rho P_g  \ket{ \psi^{(i)}_k }.
    \label{eq:partial_trace}
\end{eqnarray}
Since $P_g = \otimes_i P_g^i$ and $P_g^i \ket{ \psi^{(i)}_k } = 0 $ unless $k=0,1$, it follows that the RHS of (\ref{eq:partial_trace}) is vanishing for $k>1$. Thus we have that the reduced density matrix on the space of the 2D model  (Eq.~(\ref{eq:Reduced density  matrix on ladder})) becomes 
\begin{equation}
    \rho^{(i)} \approx \sum_{k=0}^{1} \langle \psi^{(i)}_k | \rho \ket{ \psi^{(i)}_k }=\tilde{\rho}^{(i)}.  
\end{equation}
Hence proved. 
\end{proof} 

The next corollary follows straightforwardly from \textbf{Proposition 4}.

\noindent$\diamond$ \textbf{Corollary 4.1.} \emph{The reduced density matrix $\rho_S$ for a subsystem composed of a set $S$ of arbitrary number of rungs obtained from the state $\rho$ of  the 2D model and the reduced density matrix $\tilde{\rho}_S$ of the corresponding set $S$ of spins obtained from the state $\tilde{\rho}$ of the 1D $XXZ$ model obeys $\rho_S\approx\tilde{\rho}_S$.}

\noindent\textbf{Proposition 4} and \textbf{Corollary 4.1} can be verified by computing the trace distance between $\rho_S$ and $\tilde{\rho}_S$. This implies that any function of the density matrices, for example, entanglement quantified over subsystems using their reduced density matrices, such as the bipartite entanglement between two chosen parties~\cite{horodecki2009,guhne2009}, entanglement spectrum~\cite{Li2008}, distance-based multipartite entanglement~\cite{Wei2005,Orus2008,Biswas2014}, or multiparty quantum correlation quantified by monogamy-bases approaches~\cite{ekert1991,bennett1996a,coffman2000,terhal2004,kim2012,bera2012,Rao2013,dhar2017} should also match for $\rho_S$ and $\tilde{\rho_S}$. To demonstrate this, we choose bipartite entanglement over a subsystem $S$ constituted of two nearest-neighbor rungs in the 2D lattice, which corresponds to two nearest-neighbor spins in the 1D effective model. We numerically compute the bipartite entanglement $E_{i,i+1}$, between the two rungs $i$ and $i+1$ in the 2D model from their reduced density matrix $\rho_{i,i+1}$, and the bipartite entanglement $\tilde{E}_{i,i+1}$ between the spins $(i,i+1)$ in the 1D effective $XXZ$ model from their reduced density matrix $\tilde{\rho}_{i,i+1}$, where negativity~\cite{peres1996,horodecki1996,zyczkowski1998,vidal2002,lee2000,plenio2005,leggio2020} (see Appendix~\ref{app:ent_measures} for a definition) is used for entanglement quantification. These reduced density matrices $\rho_{i,i+1}$ and $\tilde{\rho}_{i,i+1}$ are obtained respectively by tracing out all other rungs other than the rungs $(i,i+1)$ from the thermal state $\rho$ of the 2D model, and by tracing out all other spins except the spins $i$ and $i+1$ from the thermal state $\tilde{\rho}$ of the 1D model (see Sec.~\ref{subsubsec:thermal_states}). To ensure low-energy of the states, we fix $\beta J_\perp=10^2$, while $J_\perp=10^2$.

\begin{figure*}
    \includegraphics[width=0.8\textwidth]{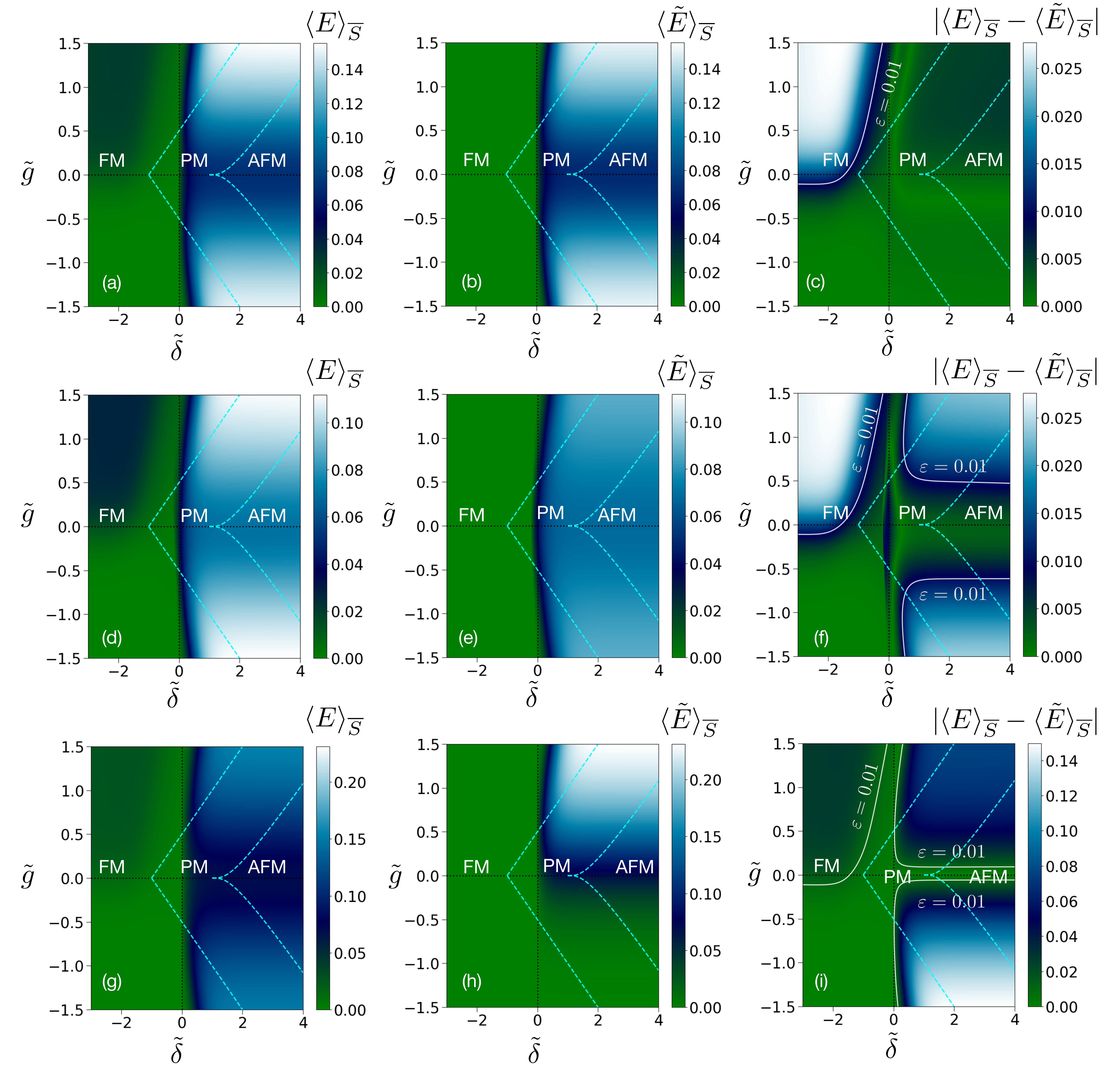}
    \caption{Variations of $\langle E\rangle_{\overline{S}}$, $\langle \tilde{E}\rangle_{\overline{S}}$, and $\varepsilon=|\langle E\rangle_{\overline{S}}-\langle \tilde{E}\rangle_{\overline{S}}|$ as functions of $\tilde{\delta}$ and $\tilde{g}$ for a $3\times 2$ lattice. The top, middle, and bottom horizontal panels are respectively for the measurement of $M^{zz}_i$, $M^{zx}_i$, and $M^{xx}_i$.  All quantities plotted are dimensionless.}
    \label{fig:average_ent}
\end{figure*}

The variations for $E_{i,i+1}$ and $\tilde{E}_{i,i+1}$ as functions of $(\tilde{\delta},\tilde{g})$ are included in Fig.~\ref{fig:entanglement_nn} in the case of a two-leg and a three-leg system.  Similar to the expectation values of observables discussed in Sec.~\ref{sec:mapping_observables}, to quantitatively test the performance of the 1D effective theory in estimating $E_{i,i+1}$, we look at the variation of $\varepsilon=|E_{i,i+1}-\tilde{E}_{i,i+1}|$ as functions of $\tilde{\delta}$ and $\tilde{g}$, and find that $\varepsilon\leq J_\perp^{-1}=10^{-2}$ for the whole of the AFM and the PM phases, while relatively high values of $\varepsilon$ occurs in the FM phases for $\tilde{g}\geq 0$. These observations remain qualitatively unchanged with a change in the value of $L$ from $2$ to $3$. To check whether these observations change with a change of the subsystem over which entanglement is computed, or with a change from the bipartite entanglement to a multipartite quantum correlation, we further consider (a) bipartite entanglement, as quantified by negativity, over a bipartition of the entire system, and (b) a monogamy-based~\cite{ekert1991,bennett1996a,coffman2000,terhal2004,kim2012,dhar2017} multiparty quantum correlation, referred to as the monogamy score~\cite{bera2012,Rao2013,dhar2017}, over the entire system, and find the observations reported so far to be valid. See Appendix~\ref{app:qc} for details. Also, note that for the entirety of the $(\tilde{\delta},\tilde{g})$ plane considered in this paper, the behaviour of different quantum correlations is qualitatively mimicked by the same for their corresponding $XXZ$ counterparts, although the perturbation theory may not be valid everywhere on the plane.

\subsection{Measurement-based quantification of entanglement}
\label{subsec:localizable_entanglement}

Given a quantum state $\rho$ of a composite quantum system, one may also quantify entanglement over a subsystem using a measurement-based protocol~\cite{divincenzo1998,verstraete2004,verstraete2004a,popp2005,sadhukhan2017}. For a bipartition $S:\overline{S}$ of the system, a local projection measurement on one of the subsystems, say, $S$, leads to an ensemble of post-measured states $\varrho_{S\overline{S}}^{(i)}=P_S^{(i)}\rho_{S\overline{S}}P_S^{(i)\dagger}/\mbox{Tr} \left[P_S^{(i)}\rho_{S\overline{S}}P_S^{(i)\dagger}\right]$ with probabilities $p_i=\mbox{Tr}\left[P_S^{(i)}\rho_{S\overline{S}}P_S^{(i)\dagger}\right]$. Here, $P_S^{(i)}=\ket{\psi_S^{(i)}}\bra{\psi_S^{(i)}}$, $i=0,2,\cdots,d_S-1$ are projection operators corresponding to the eigenkets $\ket{\psi_S^{(i)}}$ of a chosen Hermitian operator $M_S$ on the Hilbert space of $S$ with dimension $d_S$. For each choice of $M_S$, one can define average entanglement in the post-measured states of $\overline{S}$, given by 
\begin{eqnarray}
\langle E\rangle_{\overline{S}}=\sum_{i=0}^{d_S-1}p_iE(\varrho_{\overline{S}}^{(i)}),
\end{eqnarray}
where $\varrho_{\overline{S}}^{(i)}=\mbox{Tr}_S\varrho_{S\overline{S}}^{(i)}$. Note that the subsystem $\overline{S}$ can be a collection of smaller subsystems, and the entanglement measure $E$ can be either a bipartite, or a multiparty measure. In this work, we choose bipartite entanglement quantified by negativity. For a given operator $M_S$ with a low-energy description (i.e., if $P_g M_S P_g\neq 0$), we ask the  question as to \emph{whether the value $\langle E\rangle_{\overline{S}}$ obtained from a low-energy state $\rho_{S\overline{S}}$ of the 2D lattice is approximated by $\langle\tilde{E}\rangle_{\overline{S}}$ via a measurement of $\tilde{M}_S$ in the corresponding state $\tilde{\rho}_{S\overline{S}}$ of the 1D $XXZ$ model}? We ask this question for two classes of operators - one keeps the low-energy sector invariant, i.e, $P_gM_SP_g=M_S$, and the other which takes a low-energy state out of the low-energy manifold, i.e., $P_gM_SP_g\neq M_S$.

We now identify each $S$ as the rung $i$, and for demonstration, we choose three examples of single-rung operators, given by $M_i^{zz}=\sigma^z_{i,1}\otimes\sigma^z_{i,2}$ belonging to the first category, and  $M_i^{xx}=\sigma^x_{i,1}\otimes\sigma^x_{i,2}$, and $M_i^{zx}=\sigma^z_{i,1}\otimes\sigma^x_{i,2}$ belonging to the second category, with their respective low-energy components given in Table~\ref{operators}.  For these operators, we compute $\langle E\rangle_{\overline{S}}$ in a $N\times 2$ ladder, $\langle \tilde{E}\rangle_{\overline{S}}$ in the $N$-spin 1D effective model, and   $\varepsilon=|\langle E\rangle_{\overline{S}}-\langle \tilde{E}\rangle_{\overline{S}}|$ for the thermal state (see Sec.~\ref{subsubsec:thermal_states}), with $\beta J_\perp=10^2$, where $J_\perp=10^2$. In all our calculations, entanglement in the subsystem $\overline{S}$ is computed over the bipartition $1$ rung : rest of the subsystem ($1$ spin : rest of the subsystem) in the case of the 2D model (effective 1D model).  The behaviours of $\langle E\rangle_B$, $\langle\tilde{E}\rangle_B$, and $\varepsilon$, in the case of a $3\times 2$ lattice, as functions of $\tilde{\delta}$ and $\tilde{g}$, in the case of these single-rung operators are depicted in Fig.~\ref{fig:average_ent}. It is clear from the figure that the performance of the 1D $XXZ$ model as a proxy for the 2D model worsens as one shifts from the first category of operators to the second. We also point out that in the case of $M^{zz}_i$, the 1D $XXZ$ model is a good substitute for the 2D model in the entire PM and AFM phases as well as in the FM phase where $\tilde{g}<0$.

\section{Conclusion and Outlook}
\label{sec:outlook}

In this paper, we consider a spin-$1/2$ isotropic Heisenberg model in a magnetic field on a rectangular zig-zag lattice of size $N\times L$. We show that in some regimes of system parameters, irrespective of the values of $N$ and $L$, the 2D model can be well-approximated by an 1D spin-$1/2$ $XXZ$ model. In particular, we show that for specific states in the low-energy manifold of the 2D model, matrix elements of typical Hermitian operators, and non-local quantum correlations such as entanglement, the 1D $XXZ$ model provides a satisfactory proxy even when the perturbation parameters are not small. We further consider the quantification of measurement-based entanglement, where a measurement over a Hermitian operator is required, and find that the 1D model can mimic the 2D model for certain choices of the Hermitian operators. While we demonstrate these results for low values of $N$ and $L$, these positive findings open up an opportunity for investigating observables as well as entanglement in the case of $2D$ models with higher $N$ and $L$ using the 1D $XXZ$ model as a proxy.  

\begin{figure}
    \centering
    \includegraphics[width=0.8\linewidth]{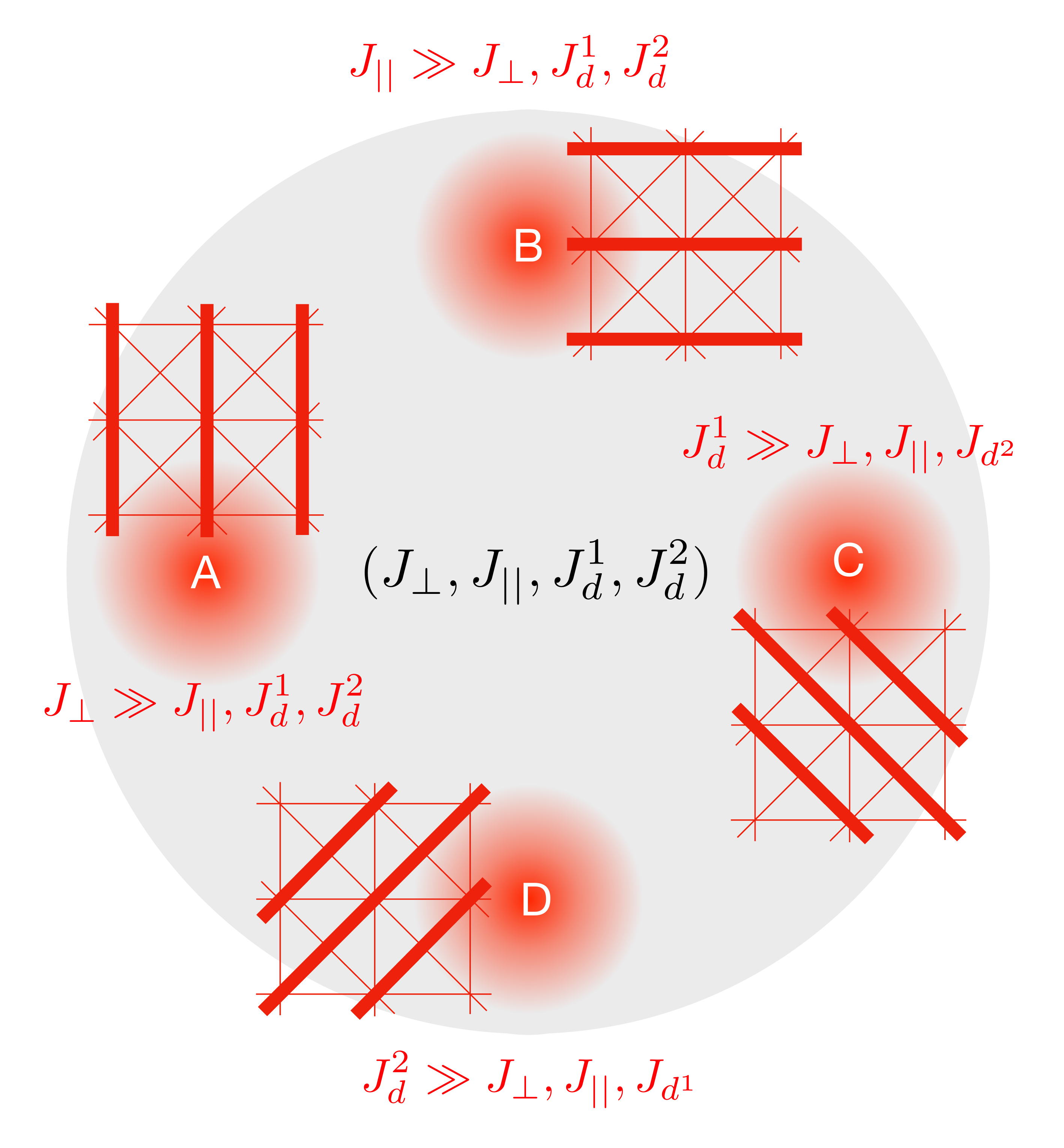}
    \caption{Schematic representation of the subspaces denoted by (A) $J_\perp\gg J_{||},J_d^1,J_d^2$, (B) $J_{||}\gg J_{\perp},J_d^1,J_d^2$, (C) $J_d^1\gg J_{\perp},J_{||},J_{d}^2$, and (D) $J_d^2\gg J_\perp,J_{||},J_{d}^1$ in the space of the parameters $J_\perp,J_{||},J_d^1$, and $J_d^2$. The thick lines in he lattices represent the strong couplings. While this paper deals with only the subspace (A), the 1D effective model in each of these subspaces can be worked out following the same methodology.}
    \label{fig:parameter_space}
\end{figure}

We conclude with a discussion on possible future works. An interesting direction would be to generalize the calculation where ground-states of individual rungs have higher degeneracies ($d>2$), so that 2D models with $d$ degrees of freedom per cite can be addressed. It is also important to ask whether expectation values of Hermitian operators defined on the 2D model for specific purposes, such as order parameters~\cite{Almeida_2008,Almeida_2008a}, or entanglement witness operators~\cite{krammer2009,Scheie2021}, can be estimated using the effective 1D model. Moreover, motivated from the results on measurement-based quantification of entanglement, other quantum correlations belonging to the quantum information-theoretic paradigms~\cite{Modi2012,Bera_2018}, such as quantum discord~\cite{Ollivier2001,Henderson_2001} and quantum work deficit~\cite{Oppenheim2002,Horodecki2005}, requiring measurements on subsystems can be investigated. While our mapping to 1D $XXZ$ model works for specific sates (see Sec.~\ref{subsec:merit_of_mapping}) in the low-energy manifold of the 2D model,  the calculation can be extended to other low-energy states by working to higher orders in perturbation theory so that all degeneracies are lifted. We expect that our formalism can be applied to the strong leg, left-diagonal, and right-diagonal limits also, providing a large subspace in the parameter space of the coupling constants of the 2D model (see Fig.~\ref{fig:parameter_space}) where the 1D effective theory works.  It is also worthwhile to note that the advantage in using the 1D effective model lies in the drastic reduction in the degrees of freedom in certain parameter regimes.   It would be interesting to look for other quantum many-body models where this happens.

\acknowledgements 

We acknowledge the use of \href{https://github.com/titaschanda/QIClib}{QIClib} -- a modern C++ library for general purpose quantum information processing and quantum computing.

\appendix

\section{Rotational symmetry of the effective Hamiltonian}

In this appendix, we take a slightly more abstract approach and argue that the effective Hamiltonian preserves the rotational symmetry in $z$ direction to any order in perturbation theory in ${1 / J_\perp}$. More precisely, writing (note that in this paper we work with only the leading order)
\begin{eqnarray}
\tilde H = \sum_{n=0}^\infty J_\perp^{-n} \tilde H^{(n)}  
\end{eqnarray}
we show below that the $[\tilde H^{(n)} , \mathcal{Z}] = 0, \forall n$. The general formalism of perturbation theory provides~\cite{Mila2011}  
\begin{eqnarray}
\tilde H^{(n)} &=& P_gH_0P_g+P_gH^\prime P_g+P_gH^\prime P_e(P_e H_0 P_e)^{-1}\nonumber\\
&\times&\sum_{n=0}^\infty \left[(P_e H^\prime P_e-EI)(P_e H_0 P_e)^{-1}\right]^n P_e H^\prime P_g. \nonumber\\ 
\end{eqnarray}
Note that the only operators on the RHS are $H'$ and $ P_g$ (recall that $P_g$ are the projector to ground state manifold). The proof that $[\tilde H^{(n)} , \mathcal{Z}] = 0$ then follows simply from the fact that $[H' , \mathcal{Z}] = 0$ and $[P_g, \mathcal{Z}] = 0$ -- the former is simply the consequence of rotational symmetry of the 2D system (Eq.~(\ref{eq:system_Hamiltonian})) and the latter we prove below. Note that we can construct $P_g$ explicitly as
\begin{eqnarray}
P_g = \sum_l \ket{\Psi_l} \bra{\Psi_l}. 
\end{eqnarray}
Since from Eq.~(\ref{eq:Action of Z on ground state manifold}), $\ket{\Psi_l}$ is an eigenvector of the Hermitian operator ${\mathcal Z}$ , it is obvious that $[P_g, \mathcal{Z}] = 0$.

\section{Low-energy effective Hamiltonian for OBC along rungs}
\label{app:obc_along_rungs_coefficients}

In this Section, we present the important steps of determining the effective coupling strengths $\tilde{J}_{xy}$, $\tilde{J}_{zz}$, and $\tilde{h}$ in Eq.~(\ref{eq:1d_general}), with $\ket{\psi_0^{(i)}}$ and $\ket{\psi_1^{(i)}}$ given in Eqs.~(\ref{eq:all_up}) and (\ref{eq:all_up_except_one_obc}). Similar to Sec.~\ref{sec:details}, we have assumed $J_\perp=1$ for brevity. Also, For ease of reference, we write 
\begin{eqnarray}
H^\prime=H_L+H_D+H_F. 
\label{eq:perturbation_Hamiltonian}
\end{eqnarray}
Here, $H_L$ is the leg-Hamiltonian given by
\begin{eqnarray}
H_L&=&\frac{J_{||}}{4}\sum_{i=1}^N\sum_{j=1}^L\vec{\sigma}_{i,j}.\vec{\sigma}_{i+1,j},
\label{eq:perturbation_Hamiltonian_leg}
\end{eqnarray}
and $H_D=H_{D_L}+H_{D_R}$ represents the diagonal interactions with
\begin{eqnarray}
H_{D_L}&=&\frac{J_{d}^1}{4}\sum_{i=1}^N\sum_{j=1}^{L}\vec{\sigma}_{i,j+1}.\vec{\sigma}_{i+1,j},
\label{eq:perturbation_Hamiltonian_diag_left}
\end{eqnarray}
and
\begin{eqnarray}
H_{D_R}&=&\frac{J_{d}^2}{4}\sum_{i=1}^N\sum_{j=1}^{L}\vec{\sigma}_{i,j}.\vec{\sigma}_{i+1,j+1}
\label{eq:perturbation_Hamiltonian_diag_right}
\end{eqnarray}
representing the \emph{left} and \emph{right} diagonals respectively (see Fig.~\ref{fig:fig1_schematics}).  The field-part $H_F$ of the perturbation Hamiltonian is  given by 
\begin{eqnarray} 
H_F&=& -\frac{\Delta h}{2}\sum_{i=1}^N\sum_{j=1}^L\sigma^z_{i,j}.
\label{eq:perturbation_Hamiltonian_field}
\end{eqnarray}
We start with the field-contribution in $H^\prime$, and write $H_F=\sum_{i=1}^NH_{F_i}$, $H_{F_i}$ being the field-term corresponding to the $i$th rung. The action of $H_{F_i}$ on the states $\ket{\psi_{0,1}^{(i)}}$ along with the definition of $\tau^z_i$ leads to the effective Hamiltonian
\begin{eqnarray}
\tilde{H}_{F} &=& -\frac{\Delta h}{2}\left[N(L-1)I+\sum_{i=1}^N\tau_i^z\right],
\label{eq:prop1_0}
\end{eqnarray}
where $I$ is the identity matrix on the ladder Hilbert space. We next consider the leg-term $H_L$, and treat the terms corresponding to the $xy$- and the $zz$-interactions separately, i.e., $H_L=H_L^{xy}+H_{L}^{zz}$. In the basis $\ket{\psi_{k_i}^{(i)}\psi_{k_{i+1}}^{(i+1)}}$, action of the $i$th term of $H_{L}^{xy}$ provides non-zero values for only $\langle\psi^{(i)}_0\psi^{(i+1)}_1|\dots|\psi^{(i)}_1\psi^{(i+1)}_0\rangle$ and its hermitian conjugate, leading to the effective Hamiltonian 
\begin{eqnarray}
\tilde{H}_{L}^{xy} 
&=& \frac{J_{||}}{4}\sum_{i=1}^N\left(\tau_i^x\tau_{i+1}^x+\tau_{i}^y\tau_{i+1}^y\right).
\label{eq:prop1_1}
\end{eqnarray}
On the other hand, $i$th term of $H_{L}^{zz}$ additionally has non-zero values for $\bra{\psi_0^{(i)}\psi_0^{(i+1)}}\dots\ket{\psi_0^{(i)}\psi_0^{(i+1)}}$ and $\bra{\psi_1^{(i)}\psi_1^{(i+1)}}\dots\ket{\psi_1^{(i)}\psi_1^{(i+1)}}$ also, which leads to 
\begin{eqnarray}
\tilde{H}_L^{zz} &=& \frac{NJ_{||}}{16}\left[3L-4+A_1\right]I\nonumber\\
&&+\frac{J_{||}}{8}\left[L-A_1\right]\sum_{i=1}^N\tau^z_i\nonumber\\
&&+\frac{J_{||}}{16}\left[4-L+A_1\right]\sum_{i=1}^N\tau_i^z\tau_{i+1}^z,
\label{eq:prop1_2}
\end{eqnarray}
with the constant $A_1$ given in Eq.~(\ref{eq:A1}).

Next, we pick up the diagonal terms $H_{D_L}$ and $H_{D_R}$ in the perturbing Hamiltonian (see Eqs.~(\ref{eq:perturbation_Hamiltonian_diag_left})-(\ref{eq:perturbation_Hamiltonian_diag_right})), and proceed by writing
$H_{D}^{xy}=H_{D_L}^{xy}+H_{D_R}^{xy}$ and $H_{D}^{zz}=H_{D_L}^{zz}+H_{D_R}^{zz}$. Calculation similar to the effective Hamiltonian of $H_L^{xy}$ and $H_{L}^{zz}$ provides
\begin{eqnarray}
\tilde{H}_{D}^{xy} 
&=& \frac{1}{4}A_0(J_d^1+J_d^2)\sum_{i=1}^N\left(\tau_i^x\tau_{i+1}^x+\tau_{i}^y\tau_{i+1}^y\right).\nonumber\\ 
\label{eq:prop1_3}
\end{eqnarray}
and 
\begin{eqnarray}
\tilde{H}_D^{zz} &=& \frac{N}{16}\left[(L-1+A_3+2A_2)(J_d^1+J_d^2)\right]I\nonumber\\
&&+\frac{1}{8}(L-1-A_3)(J_d^1+J_d^2)\sum_{i=1}^N\tau^z_i\nonumber\\
&&+\frac{1}{16}\left[(L-1+A_3-2A_2)(J_d^1+J_d^2)\right]\sum_{i=1}^N\tau_i^z\tau_{i+1}^z,\nonumber\\
\label{eq:prop1_4}
\end{eqnarray}
with $A_0$, $A_2$, and $A_3$ given by Eqs.~(\ref{eq:A0}), (\ref{eq:A2}) and (\ref{eq:A3}) respectively. 
Therefore, combining Eqs.~(\ref{eq:prop1_0}), (\ref{eq:prop1_1}), (\ref{eq:prop1_2}), (\ref{eq:prop1_3}), and (\ref{eq:prop1_4}), the effective Hamiltonian
\begin{eqnarray}
\tilde{H}&=&\tilde{H}_F+ \tilde{H}^{xy}_L+\tilde{H}^{zz}_L+\tilde{H}^{xy}_D+\tilde{H}^{zz}_D
\end{eqnarray} 
turns out to be as given in Eq.~(\ref{eq:1d_general}), with the effective couplings and the additive constant are given in Eqs.~(\ref{eq:effective_coupling_open_closed}).

In the following, we include the details of the above mapping in the case of $L=2$~\cite{Tonegawa1998,Mila1998,Tandon1999,totsuka1998,Batchelor2003,Batchelor2003a}, $L=3$~\cite{kawano1997,Tandon1999}, and $L=4$, and provide the expressions for  $\tilde{J}_{xy}$, $\tilde{J}_{zz}$, and $\tilde{h}$, where $J_\perp$ is assumed to be $1$.  

\subsection{\texorpdfstring{$L=2$}{L=2}}

For $L=2$~\cite{Tonegawa1998,Mila1998,Tandon1999,totsuka1998,Batchelor2003,Batchelor2003a}, The $i$th rung has doubly degenerate ground states ($d=2$), given by 
\begin{eqnarray}
\ket{\psi_0^{(i)}} &=& \ket{00},\nonumber \\
\ket{\psi_1^{(i)}} &=& \frac{1}{\sqrt{2}}(\ket{01}-\ket{10}),
\label{eq:standard_basis}
\end{eqnarray}
at $h^\prime=1$, with the ground-state energy $E_0=- 3\perp/4$. The coupling constants in $\tilde{H}$ are given by 
\begin{eqnarray}
\label{eq:two_leg_jxy}
\tilde{J}_{xy} &=& \left[2J_{||}-J_d^1-J_d^2\right]/8, \\
\label{eq:two_leg_jz}
\tilde{J}_{zz} &=&\left[2 J_{||}+J_d^1+J_d^2\right]/16,  \\
\label{eq:two_leg_h}
\tilde{h} &=& \left[2J_{||}-4\Delta h+J_{d}^{1}+J_{d}^{2}\right]/8.
\end{eqnarray}
where we have  used PBC along the legs.
Using OBC along the legs changes the 1D LEH to Eq.~(\ref{eq:1d_general_obc}), 
with $\tilde{J}_{xy}$, $\tilde{J}_{zz}$, and $\tilde{h}$ given by Eqs.~(\ref{eq:two_leg_jxy})-(\ref{eq:two_leg_h}), and 
\begin{eqnarray}
\label{eq:two_leg_h_obc}
\tilde{h}^\prime &=&-\left[2J_{||}+J_d^1+J_d^2\right]/16. 
\end{eqnarray}

\subsection{\texorpdfstring{$L=3$}{L=3}}

For $L=3$ in Eq.~(\ref{eq:system_Hamiltonian})~\cite{kawano1997,Tandon1999}, 
\begin{eqnarray}
\ket{\psi_0^{(i)}} &=& \ket{000}, \nonumber \\ 
\ket{\psi_1^{(i)}} &=&(\ket{001}-2\ket{010}+\ket{100})/\sqrt{6},
\end{eqnarray} 
which are degenerate at $h^\prime=3/2$, with ground state energy $E_0=-7/4$. The effective coupling constants in Eq.~(\ref{eq:1d_general}) are given by    
\begin{eqnarray}
\tilde{J}_{xy} &=&\left[3J_{||}-2(J_d^1+J_d^2)\right]/12,\\
\tilde{J}_{zz}&=&\left[9J_{||}+4(J_d^1+J_d^2)\right]/72,\\
\tilde{h} &=&\left[9J_{||}+11(J_d^1+J_d^2)-18\Delta h\right]/36.
\label{eq:effective_constants_pbc_three}
\end{eqnarray}
In the case of OBC along the legs, the effective Hamiltonian is modified to Eq.~(\ref{eq:1d_general_obc}) with 
\begin{eqnarray}
\tilde{h}^\prime &=&-\left[9J_{||}+11(J_d^1+J_d^2)\right]/72.
\end{eqnarray}

\subsection{\texorpdfstring{$L=4$}{L=4}}

For $L=4$, the doubly-degenerate ground states at $h^\prime=1+1/\sqrt{2}$ are given by
\begin{eqnarray}
\ket{\psi_0} &=& \ket{0000}, \nonumber \\ 
\ket{\psi_1} &=&\Big[-\ket{0001}+a\ket{0010}-a\ket{0100}\nonumber\\
&&+\ket{1000}\Big]/\sqrt{2+2a^2},
\end{eqnarray}
with $a=1+\sqrt{2}$. The effective coupling constants in Eq.~(\ref{eq:1d_general}) are given by
\begin{eqnarray}
\tilde{J}_{xy} &=&\left[8J_{||}-(2+3\sqrt{2})(J_d^1+J_d^2)\right]/32, \nonumber \\
\tilde{J}_{zz}&=&\big[12J_{||}+(2\sqrt{2}+5)(J_d^1+J_d^2)\big]/128, \nonumber\\
\tilde{h} &=&\big[20J_{||}-32\Delta h+(2\sqrt{2}+19)(J_d^1+J_d^2)\big]/64.\nonumber\\ 
\label{eq:effective_constants_pbc_four}
\end{eqnarray}

\section{Trace distance}
\label{app:distance}

Here we briefly discuss the trace and Bures distance metrics used in this paper for quantifying the distance between density matrices of subsystems of the quantum spin ladder. Given two density operators $\rho_1$ and $\rho_2$ defined on a Hilbert space of dimension $d$, the trace-distance between $\rho_1$ and $\rho_2$ is defined as~\cite{wilde_book} 
\begin{eqnarray}
d_T(\rho_1,\rho_2) &=& \frac{1}{2}\sqrt{(\rho_1-\rho_2)^\dagger(\rho_1-\rho_2)} \nonumber\\ 
&=& \frac{1}{2}\sum_{i=1}^d|\lambda_i|,
\end{eqnarray}
where $\lambda_i$ are the singular values of the Hermitian matrix $\rho_1-\rho_2$. Note that $\rho_1-\rho_2$ is not necessarily positive.

\section{Odd number of spins on a rung with vanishing magnetic field}
\label{app:zero_field}

For ease of presenting the steps of the calculation, we set $j_d^1=J_d^2=0$, and redefine the states in Eqs.~(\ref{eq:0state0}) and (\ref{eq:0state1}) as
\begin{eqnarray}
\ket{\psi_0^{(i)}}&=&\sum_{\{j\}\in S}a_{\{j\}}\ket{\{j\}}, \nonumber \\
\ket{\psi_1^{(i)}}&=&\sum_{\{j\}\in S}a_{\{j\}}\ket{\{\overline{j}\}}.
\label{eq:new_state}
\end{eqnarray}
Here, $\{j\}=\{j_1,j_2,\cdots,j_l\}$, such that $j_1<j_2<\dots<j_{(l-1)}<j_l$, with each $j\in S$ where $S\equiv\{1,2,\dots,L-1,L\}$. In each of the terms in $\ket{\psi_0^{(i)}}$ $(\ket{\psi_1^{(i)}})$, the indices $\{j\}$ indicate the lattice sites of the spins that are in the state $\ket{1}$ ($\ket{0}$), and  $\ket{\{\overline{j}\}}$ is the complement of $\ket{\{j\}}$, such that $\ket{\psi_1^{(i)}}=\left(\sigma^{x}\right)^{\otimes L}\ket{\psi_0^{(i)}}$ (see Eq.~(\ref{eq:0state1})).  
We follow the prescription given in Sec.~\ref{subsec:leh}. First, we write the leg term as $H_L=H_{L}^{xy}+H_{L}^{zz}$ (see Appendix~\ref{app:obc_along_rungs_coefficients}), where the action of the $i$th term in $H_{L}^{xy}$, corresponding to the $i$th rung, on two neighbouring rungs $(i,i+1)$ is given by
\begin{eqnarray}
&&\bra{\psi_1^{(i)}\psi_0^{(i+1)}}H_{L_i}^{xy}\ket{\psi_0^{(i)}\psi_1^{(i+1)}}\nonumber\\
&=& \sum_{\{j\}\in S}\sum_{\{j^\prime\} \in S}\sum_{k}a_{\{j\}}a_{\{j^\prime\}}a_{\{p\}}a_{\{q\}},\nonumber \\
&&\bra{\psi_0^{(i)}\psi_1^{(i+1)}}H_{L_i}^{xy}\ket{\psi_1^{(i)}\psi_0^{(i+1)}}\nonumber\\ &=& \sum_{\{j\}\in S}\sum_{\{j^\prime\}\in S}\sum_{k}a_{\{j\}}a_{\{j^\prime\}}a_{\{p\}}a_{\{q\}},
\label{eq:dynamical_term_L_leg_h0}
\end{eqnarray}
with $\{p\}=\{p_1,p_2,\cdots,p
_l\} \in S_1$, where $S_1=S\backslash\{k,\{j\}\}$ is the set of all elements in $S$ except $k$ and $\{j\}$. Similarly, $\{q\}=\{q_1,q_2,..q_l\} \in S_2$ with  $S_2=S\backslash\{k,\{j^\prime\}\}$, and $k\in S_3$ with $S_3=S\backslash\{\{j\},\{j^\prime\}\}$. The action of $H_{L_i}^{xy}$ on $\ket{\psi_0^{(i)}\psi_{0}^{(i+1)}}$ and $\ket{\psi_1^{(i)}\psi_{1}^{(i+1)}}$ takes the system out of the ground state manifold.  The effective Hamiltonian due to $H_{L}^{xy}$ therefore becomes
\begin{eqnarray}
\tilde{H}_{L}^{xy} 
&=&A \frac{J_{||}}{4}\sum_{i=1}^N\left(\tau_i^x\tau_{i+1}^x+\tau_{i}^y\tau_{i+1}^y\right),
\end{eqnarray}
where 
\begin{eqnarray}
A=\sum_{\{j\}\in S}\sum_{\{j^\prime\}\in S}a_{\{j\}}a_{\{j^\prime\}}a_{\{p\}}a_{\{q\}}.
\label{eq:a4}
\end{eqnarray} 
On the other hand, non-zero matrix elements of $H_{L_i}^{zz}$ corresponding to the neighbouring rungs $(i,i+1)$ are 
\begin{eqnarray}
&&\bra{\psi_0^{(i)}\psi_0^{(i+1)}}H_{L_i}^{zz}\ket{\psi_0^{(i)}\psi_0^{(i+1)}}\nonumber\\ &=& \sum_{\{j\}\in S}\sum_{\{j^\prime\}\in S}\left[2(n_1+n_2)-L\right]a^2_{\{j\}}a^2_{\{j^\prime\}},\nonumber \\
&&\bra{\psi_0^{(i)}\psi_1^{(i+1)}}H_{L_i}^{zz}\ket{\psi_0^{(i)}\psi_1^{(i+1)}}\nonumber\\ &=& \sum_{\{j\}\in S}\sum_{\{j^\prime\}\in S}\left[L-2(n_1+n_2)\right]a^2_{\{j\}}a^2_{\{j^\prime\}},\nonumber \\
&&\bra{\psi_1^{(i)}\psi_0^{(i+1)}}H_{L_i}^{zz}\ket{\psi_1^{(i)}\psi_0^{(i+1)}}\nonumber\\ &=& \sum_{\{j\}\in S}\sum_{\{j^\prime\}\in S}\left[L-2(n_1+n_2)\right]a^2_{\{j\}}a^2_{\{j^\prime\}},\nonumber \\
&&\bra{\psi_1^{(i)}\psi_1^{(i+1)}}H_{L_i}^{zz}\ket{\psi_1^{(i)}\psi_1^{(i+1)}}\nonumber\\ &=&\sum_{\{j\}\in S}\sum_{\{j^\prime\}\in S}\left[2(n_1+n_2)-L\right]a^2_{\{j\}}a^2_{\{j^\prime\}},
\label{eq:static_term_L_leg_h0}
\end{eqnarray}
where $n_1$ is the cardinality of $\{j\}\cap\{j^\prime\}$, and $n_2$ is the cardinality of $(S\backslash\{j\})\cap(S\backslash{j^\prime})$. The effective Hamiltonian due to $H_{L}^{zz}$, therefore, is \small 
\begin{eqnarray}
\tilde{H}_{L}^{zz} 
&=&B \frac{J_{||}}{4}\sum_{i=1}^N\left(\tau_i^z\tau_{i+1}^z\right),
\end{eqnarray}\normalsize 
where 
\begin{eqnarray} 
B=\sum_{\{j\}\in S}\sum_{\{j^\prime\}\in S}\left[2(n_1+n_2)-L\right]a^2_{\{j\}}a^2_{\{j^\prime\}}.
\label{eq:b4}
\end{eqnarray} 
The effective Hamiltonian $\tilde{H}=\tilde{H}^{xy}_L+\tilde{H}^{zz}_L$, therefore, takes the form of an 1D XXZ Hamiltonian given by Eq.~(\ref{eq:1d_leh_zero_field}), where the coupling constants $\tilde{J}_{xy}$ and $\tilde{J}_{zz}$ are given by 
\begin{eqnarray}
\tilde{J}_{xy}&=&(J_{||}A_4)/4,\quad 
\tilde{J}_{zz}=(J_{||}B_4)/4.
\label{eq:coefficients_h0}
\end{eqnarray}

\begin{figure*}
    \includegraphics[width=0.8\textwidth]{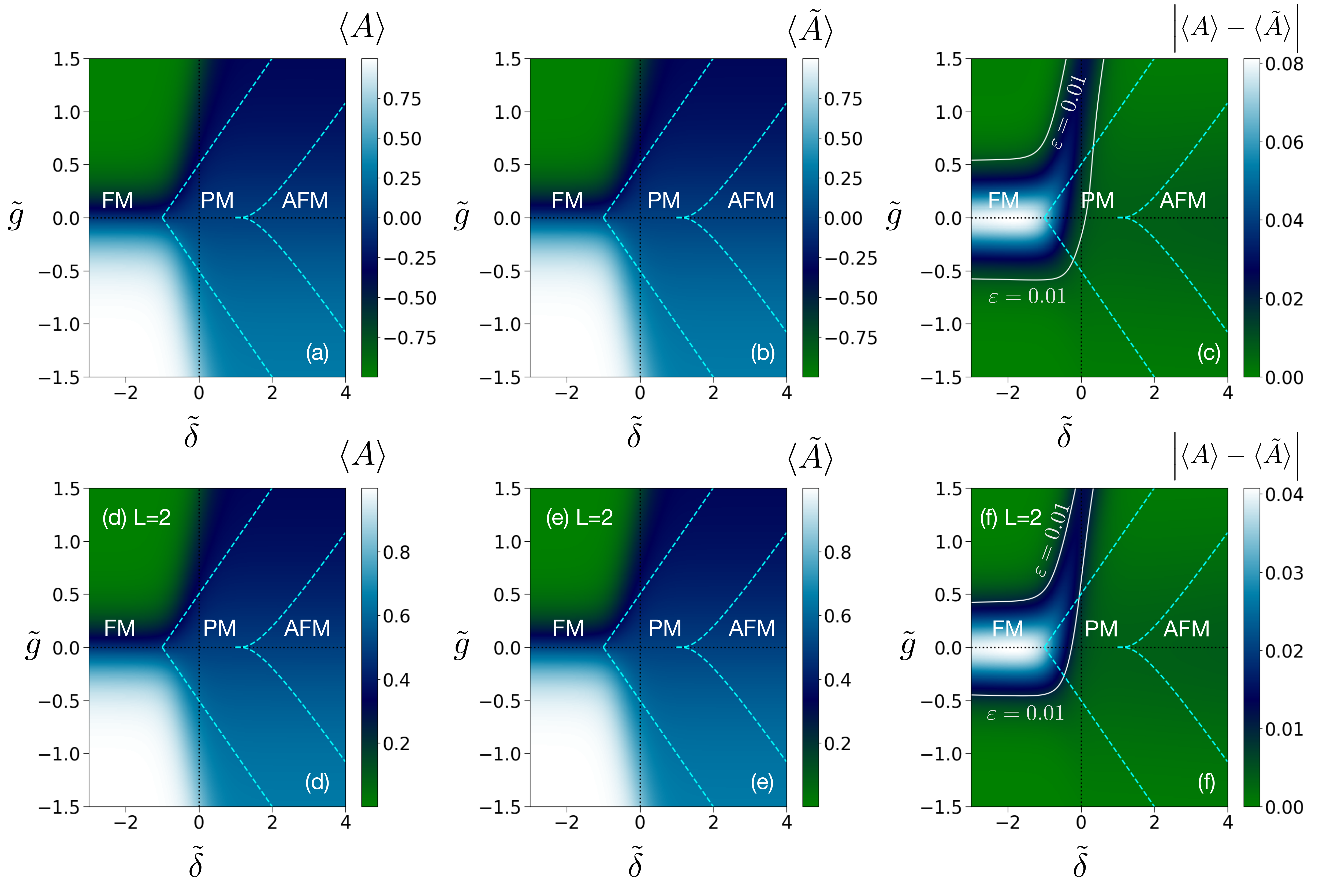}
    \caption{Variations of $\langle A\rangle$, $\langle \tilde{A}\rangle$, and $\varepsilon=\langle A\rangle-\langle \tilde{A}\rangle$ as functions of $\tilde{\delta}$ and $\tilde{g}$ for a $3\times 2$ lattice, where $A=\sigma^z_{i,1}\otimes\sigma^z_{i,2}$ (plotted in (a)), and $A=\sigma^z_{i,1}$ (plotted in (d)). The corresponding low-energy operators are $\tilde{A}=\tau^z_i$ (plotted in (b)) and $\tilde{A}=(I_i+\tau^z_i)/2$ (plotted in (e)) respectively (see Table~\ref{operators}).  All quantities plotted are dimensionless.}
    \label{fig:observables_zz}
\end{figure*}

We now demonstrate the crucial steps of the calculation in the case of three-leg ladder at $h=0$.  We first write $\ket{\psi_0}$ and $\ket{\psi_1}$ as (see Eq.~(\ref{eq:new_state}))
\begin{eqnarray}
\ket{\psi_0} &=&(a_1\ket{100}+a_2\ket{010}+a_3\ket{001}),\nonumber \\
\ket{\psi_1} &=&(a_1\ket{011}+a_2\ket{101}+a_3\ket{110}),
\end{eqnarray}
where $a_1=a_3=1/\sqrt{6}$, and $a_2=-2/\sqrt{6}$. In this notation, 
consider the action of $H_{L}^{xy}$ on one of the terms in $\ket{\psi_0^{(i)}\psi_1^{(i+1)}}$,  say, $H_{L}^{xy}(a_1^2\ket{100011})$. Here $i_1=1$, $j_1=1$, $S\equiv\{1,2,3\}$, and $S\backslash\{i_1,j_1\} \equiv {2,3}$, implying $k\in\{2,3\}$. For $k=2$, $p_1=S\backslash\{k,i_1\}=3$, $q_1=S\backslash\{k,j_1\}=3$, and for $k=3$, $p_1=S\backslash\{k,i_1\}=2$, $q_1=S\backslash\{k,j_1\}=2$. This leads to 
\begin{eqnarray}
&&\bra{\psi_0^{(i)}\psi_1^{(i+1)}}H_{L}^{xy}(a_1^2\ket{100011})\nonumber\\
&=&\sum_k a_{i_1}a_{j_1}a_{p_1}a_{q_1}=a_1^2a_2^2+a_1^2a_3^2.  
\end{eqnarray}
Proceeding in the same way for other terms in $\ket{\psi_0^{(i)}\psi_1^{(i+1)}}$, as well as the other elements of the basis 
\begin{eqnarray}
\{\ket{\psi_0^{(i)}\psi_0^{(i+1)}},\ket{\psi_0^{(i)}\psi_1^{(i+1)}},\ket{\psi_1^{(i)}\psi_0^{(i+1)}},\ket{\psi_1^{(i)}\psi_1^{(i+1)}}\},\nonumber\\
\end{eqnarray}
and substituting values of $a_1$, $a_2$, $a_3$, we obtain
\begin{eqnarray}
\bra{\psi_1^{(i)}\psi_0^{(i+1)}}H_{L}^{xy}\ket{\psi_0^{(i)}\psi_1^{(i+1)}}&=&1,\nonumber\\ 
\bra{\psi_0^{(i)}\psi_1^{(i+1)}}H_{L}^{xy}\ket{\psi_1^{(i)}\psi_0^{(i+1)}}&=&1,
\end{eqnarray}
while the other matrix elements vanish. 

To see the action of $H_{L}^{zz}$ on a pair of neighbouring rungs $(i,i+1)$, first consider
the action of $H_{L}^{zz}$ on one of the terms of in $\ket{\psi_0^{(i)}\psi_0^{(i+1)}}$, say $H^{zz}_L(a_1a_2\ket{100010})$. To determine the contribution of this term to $\bra{\psi_0^{(i)}\psi_0^{(i+1)}}H_{L_i}^{zz}\ket{\psi_0^{(i)}\psi_0^{(i+1)}}$, note that  $i_1=1$, $j_1=2$, $\mathcal{I}\equiv\{1\}$, and $\mathcal{J}\equiv\{2\}$, which leads to $n_1=0$, and $n_2=1$. Therefore, for $i_1=1$ and $j_1=2$, $n_1+n_2-(L-n_1-n_2)=-1$. Using the same procedure for other terms in $\ket{\psi_0^{(i)}\psi_0^{(i+1)}}$ as well as the other elements of the basis, and substituting values for $a_1$, $a_2$ and $a_3$, we obtain non-zero matrix elements of $H^{zz}_L$ as 
\begin{eqnarray}
\bra{\psi_0^{(i)}\psi_0^{(i+1)}}H_{L_i}^{zz}\ket{\psi_0^{(i)}\psi_0^{(i+1)}} &=& 1 \nonumber \\
\bra{\psi_0^{(i)}\psi_1^{(i+1)}}H_{L_i}^{zz}\ket{\psi_0^{(i)}\psi_1^{(i+1)}} &=& -1 \nonumber \\
\bra{\psi_1^{(i)}\psi_0^{(i+1)}}H_{L_i}^{zz}\ket{\psi_1^{(i)}\psi_0^{(i+1)}} &=& -1 \nonumber \\
\bra{\psi_1^{(i)}\psi_1^{(i+1)}}H_{L_i}^{zz}\ket{\psi_1^{(i)}\psi_1^{(i+1)}} &=& 1.  
\end{eqnarray}
This leads to $A_4=1=B_4=1$ (see Eqs.~(\ref{eq:a4}) and (\ref{eq:b4})). Therefore, the coefficients of the 1D LEH (Eq.~(\ref{eq:1d_leh_zero_field})), given by Eq.~(\ref{eq:coefficients_h0}), are
\begin{eqnarray}
\tilde{J}_{xy}&=&(J_{||})/4,\quad 
\tilde{J}_{zz}=(J_{||})/4.
\end{eqnarray}
We point out here that the coupling constants can also be determined in a similar fashion with perturbations $J_d^1,J_d^2\neq 0$, where the overall form of Eq.~(\ref{eq:1d_leh_zero_field}) remains unchanged. 


\begin{figure*}
    \includegraphics[width=0.8\textwidth]{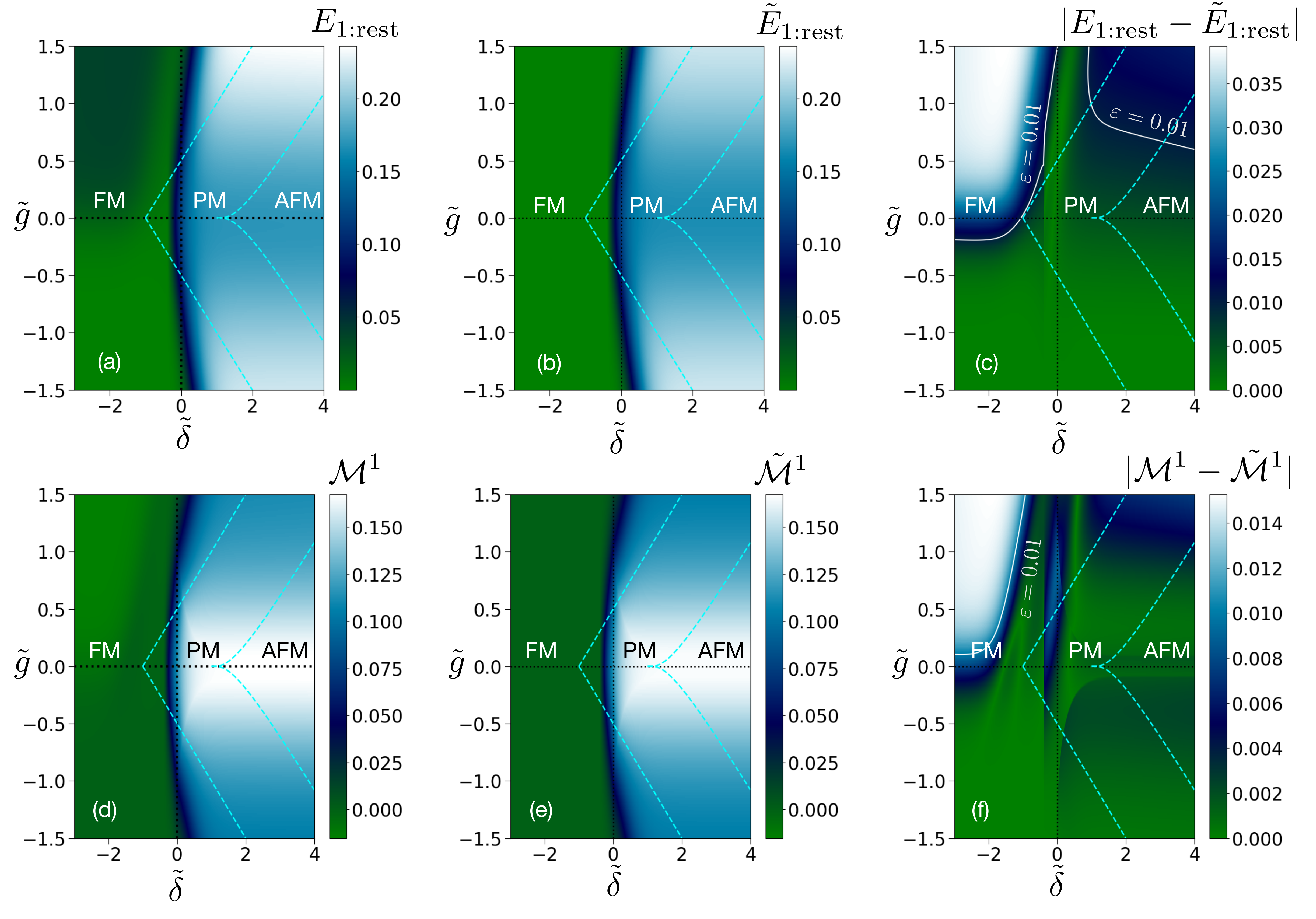}
    \caption{Variations of (a) $E_{1:\text{rest}}$, (b) $\tilde{E}_{1:\text{rest}}$, (c) $\varepsilon=|E_{1:\text{rest}}-\tilde{E}_{1:\text{rest}}|$, (d) $\mathcal{M}^1$, (e) $\tilde{\mathcal{M}}^1$, and (f) $\varepsilon=|\mathcal{M}-\tilde{\mathcal{M}}^1|$ as functions of $\tilde{\delta}$ and $\tilde{g}$ for a $3\times 2$ lattice. The continuous lines signify the value $10^{-2}$ for the respective quantities.  All quantities plotted are dimensionless.}
    \label{fig:qc}
\end{figure*}

\section{Entanglement and related measures of quantum correlations}
\label{app:ent_measures}

In this section, we provide brief definitions of the quantum correlation measures investigated in the quantum spin ladder. More specifically, we focus on bipartite entanglement~\cite{horodecki2009,guhne2009} between different parts of the system, as quantified by 
negativity~\cite{peres1996,horodecki1996,zyczkowski1998,vidal2002,lee2000,plenio2005,horodecki2009}, and multipartite quantum correlation over the complete system, or a chosen subsystem, as quantified by the entanglement monogamy score~\cite{coffman2000,terhal2004,bera2012}. 



\subsection{Negativity}
\label{app:negativity}

Entanglement between two partitions $A$ and $B$ of a bipartite quantum state $\rho_{AB}$ is quantified by a bipartite entanglement measure~\cite{horodecki2009,guhne2009}, such as the negativity~\cite{peres1996,horodecki1996,zyczkowski1998,vidal2002,lee2000,plenio2005,leggio2020}. It is defined as  
\begin{eqnarray} 
E_{S:\overline{S}}=||\rho_{S\overline{S}}^{T_{\overline{S}}}||-1,
\label{eq:neg_def}
\end{eqnarray}
with $||\varrho|| = \mbox{Tr}\sqrt{\varrho^{\dagger}\varrho}$ being the trace norm of $\varrho$ (see also Appendix~\ref{app:distance}), and $\rho_{S\overline{S}}^{T_{\overline{S}}}$ is obtained by performing partial transposition of the density matrix $\rho_{S\overline{S}}$ with respect to the subsystem  $\overline{S}$.

\subsection{Monogamy of entanglement}
\label{app:monogamy}

For a given bipartite entanglement measure $E$ and an $N$-party quantum state $\rho_{S_1S_2\cdots S_N}$, the quantum state is said to be \emph{monogamous}~\cite{ekert1991,bennett1996a,coffman2000,terhal2004,kim2012,dhar2017} for the entanglement measure $E$ if
\begin{eqnarray}
E_{S_i:\text{rest}}\geq \sum_{i\neq j}E_{S_iS_j} 
\end{eqnarray}
for $i,j\in\{1,2,\cdots,N\}$ with $S_i$ as the \emph{nodal observer}, where $E_{S_i:\text{rest}}=E(\rho_{S_i:\text{rest}})$ computed over the bipartition $S_i:\text{rest}$ of the $N$-party system, and $E_{S_iS_j}=E(\rho_{S_iS_j})$ with $\rho_{S_iS_j}=\text{Tr}_{\{S_k;k\neq i,j\}}[\rho_{S_1S_2\cdots S_N}]$. The corresponding monogamy score~\cite{bera2012,dhar2017} is defined as
\begin{eqnarray}
\mathcal{M}^{S_i} =E_{S_i:\text{rest}}-\sum_{i\neq j}E_{S_iS_j},
\end{eqnarray}
where a positive (negative) value of $\mathcal{M}^{S_i}$ indicates that the state $\rho_{S_1S_2\cdots S_N}$ is monogamous (non-monogamous) for the bipartite entanglement measure $E$ with $S_i$ as the node. Entanglement monogamy score captures multiparty quantum correlation present in the $N$-party system~\cite{Rao2013}.

\section{Expectation values of observables on a rung}
\label{app:observables}

Here we briefly discuss the variations of the expectation values of $A=\sigma^z_{i,1}$ and $A=\sigma^z_{i,1}\otimes\sigma^z_{i,2}$ on the $i$th rung of a $3\times 2$ lattice, and their corresponding low-energy components $\tilde{A}$, as functions of $\tilde{\delta}$ and $\tilde{g}$ (see Fig.~\ref{fig:observables_zz}). Note that in the strong rung-coupling limit, a rung is mapped to an effective two-level system constituting the 1D effective $XXZ$ model. Therefore, we expect  $A=\sigma^z_{i,j}$ and $A=\sigma^z_{i,1}\otimes\sigma^z_{i,2}$ to qualitatively have similar behaviour as functions of system parameters. This is confirmed by Figs.~\ref{fig:observables_zz}(a) and (d). The low-energy components of $A=\sigma^z_{i,1}\otimes\sigma^z_{i,2}$ and $A=\sigma^z_{i,j}$ are given by $\tilde{A}=\tau^z_i$ and $\tilde{A}=(I_i+\tau^z_i)/2$ respectively, and $\langle\tilde{A}\rangle$ in $\tilde{\rho}$ also match for these two observables over the entire $(\tilde{\delta},\tilde{g})$ plane. Moreover, $\varepsilon=|\langle A\rangle-\langle\tilde{A}\rangle|$ also match for both observables.

\section{Bipartite entanglement and entanglement monogamy score}
\label{app:qc}

We now discuss two specific types of  quantum correlations on a $3\times 2$ lattice -- (a) the bipartite entanglement, $E_{1:\text{rest}}$, as quantified by negativity (see Appendix~\ref{app:ent_measures}), between the rung $i=1$ and the rest of the system, and (b) the entanglement monogamy score $\mathcal{M}^1$ on the lattice, taking one of the rungs, say, rung $i=1$, as the nodal observer. Correspondingly, in the 1D effective $XXZ$ model, we compute (a) the bipartite entanglement $\tilde{E}_{1:\text{rest}}$ between the spin $i=1$ and the rest of the system, and (b) the entanglement monogamy score  $\tilde{\mathcal{M}}^1$ over the $3$-spin effective model using the spin $i=1$ as the nodal observer. Fig.~\ref{fig:qc} depicts the variations of these quantum correlations as functions of $\tilde{\delta}$ and $\tilde{g}$. Similar to the case of $E_{i,i+1}$ and $\tilde{E}_{i,i+1}$ as described in Sec.~\ref{sec:entanglement}, the qualitative variations of these quantum correlations in the 2D model and their counterparts in the 1D effective $XXZ$ model are found to be matching. Moreover, the plots of $\varepsilon=|E_{1:\text{rest}}-\tilde{E}_{1:\text{rest}}|$ and $\varepsilon=|\mathcal{M}^1-\tilde{\mathcal{M}}^1|$ as functions of $(\tilde{\delta},\tilde{g})$ are also qualitatively similar, with relatively high value of $\varepsilon$ found in the FM phase with positive magnetic field. These observations are in agreement with the ones made in the case of nearest-neighbor entanglement in the 2D and the effective 1D model in Sec.~\ref{sec:entanglement}.   

\bibliography{ref_ladder}{}
 
\end{document}